\documentclass{aa}    
\usepackage{epsfig,graphicx,color}  
\usepackage{ctable}  
\usepackage{txfonts}

\newcommand{\av}{A$_{\rm V}$}  
   
\begin{document}   
  
  \title{\emph{Herschel}'s view of the large-scale structure in the \object{Chamaeleon} dark clouds  
  \thanks{\emph{Herschel} is an ESA space observatory with science instruments provided by the European-led Principal Investigator consortia and with important participation from NASA.}}  
  
   \author{C. Alves de Oliveira\inst{1} 		
          \and N. Schneider \inst{2,3}		
          \and B. Mer\'in \inst{1}			
          \and T. Prusti \inst{4}			
          \and \'A. Ribas \inst{1,5,6}		
          \and N. L. J. Cox \inst{7}			
          \and R. Vavrek \inst{1}			
          \and V. K\"onyves\inst{8,9}		
	  \and D. Arzoumanian \inst{9}		 
      	  \and E. Puga \inst{1,10}			
          \and G. L. Pilbratt \inst{4}			 
          \and \'A. K\'osp\'al \inst{4}		 
          \and Ph. Andr\'e \inst{8}			 
          \and P. Didelon \inst{8}			 
          \and A. Men'shchikov\inst{8}		 
          \and P. Royer \inst{7}			 
          \and C. Waelkens \inst{7}			 
          \and S. Bontemps \inst{2,3}		 
          \and E. Winston \inst{4}			 
          \and L. Spezzi \inst{11}			 
          }  
  
\institute{European Space Agency (ESA/ESAC, SRE-O), P.O. Box 78, 28691 Villanueva de la Ca\~nada (Madrid), Spain \\
   \email{calves@sciops.esa.int}  
   \and  
   Universit\'{e} de Bordeaux, Laboratoire d'Astrophysique de Bordeaux, CNRS/INSU, 33270 Floirac, France
   \and   
   CNRS, LAB, UMR 5804, 33270, Floirac, France		
   \and  
   European Space Agency (ESA/ESTEC, SRE-S), Keplerlaan 1, 2201 AZ Noordwijk, The Netherlands
   \and  
   Ingenier\'ia y Servicios Aeroespaciales, European Space Agency (ESA/ESAC, SRE-O), P.O. Box 78, 28691 Villanueva de la Ca\~nada (Madrid), Spain			  
   \and  
   Centro de Astrobiologia (INTA-CSIC), P.O. Box 78, 28691 Villanueva de la Ca\~nada (Madrid), Spain	
   \and  
  Instituut voor Sterrenkunde, KU Leuven, Celestijnenlaan 200D, bus 2401, 3001 Leuven, Belgium 
   \and  
   Laboratoire AIM, CEA/DSM-CNRS-Universit\'{e} Paris Diderot, IRFU/SAp, CEA Saclay, Orme des Merisiers, 91191 Gif-sur-Yvette, France
    \and  
   IAS, CNRS (UMR 8617), Universit\'{e} Paris-Sud 11, B\^{a}timent 121, 91400 Orsay, France 
    \and  
   Vega, European Space Agency (ESA/ESAC, SRE-O), P.O. Box 78, 28691, Villanueva de la Ca\~nada (Madrid), Spain       
   \and	  
	European Southern Observatory (ESO), Karl-Schwarzschild-Strasse 2, 85748 Garching bei M\"unchen, Germany 
	}  
  
   \date{Received 24 January, 2014; accepted 23 April, 2014}  
  
  \abstract  
{The Chamaeleon molecular cloud complex is one of the nearest star-forming sites encompassing three molecular clouds (Cha~I, II, and III) with a different star-formation history, from quiescent (Cha~III) to actively forming stars (Cha~II), and reaching the end of star-formation (Cha~I).}  
{We aim at characterising the large-scale structure of the three sub-regions of the Chamaeleon molecular cloud complex by analysing new far-infrared images taken with the \emph{Herschel} Space Observatory.}  
{We derived column density and temperature maps using PACS and SPIRE observations from the \emph{Herschel} Gould Belt Survey, and applied several tools, such as filament tracing, power-spectra, $\Delta$-variance, and probability distribution functions of column density (PDFs), to derive physical properties.}  
{The column density maps reveal a different morphological appearance for the three clouds, with a ridge-like structure for Cha~I, a clump-dominated regime for Cha~II, and an intricate filamentary network for Cha~III. The filament width is measured to be around 0.12$\pm$0.04 pc in the three clouds, and the filaments found to be gravitationally unstable in Cha~I and II, but mostly subcritical in Cha~III. Faint filaments (\emph{striations}) are prominent in Cha~I showing a preferred alignment with the large-scale magnetic field. The PDFs of all regions show a lognormal distribution at low column densities. For higher densities, the PDF of Cha~I shows a turnover indicative of an extended higher density component, culminating with a power-law tail. Cha~II shows a power-law tail with a slope characteristic of gravity. The PDF of Cha~III can be best fit by a single lognormal.} 
{The turbulence properties of the three regions are found to be similar, pointing towards a scenario where the clouds are impacted by large-scale processes. The magnetic field could possibly play an important role for the star-formation efficiency in the Chamaeleon clouds if proven that it can effectively channel material on Cha~I, and possibly Cha~II, but  probably less efficiently on the quiescent Cha~III cloud.}  
  
\maketitle  
  
\section{Introduction} \label{intro}  
  
\begin{figure*}[ht]  
\begin{center}   
\hspace{0.0cm}\includegraphics[angle=0,width=18cm]{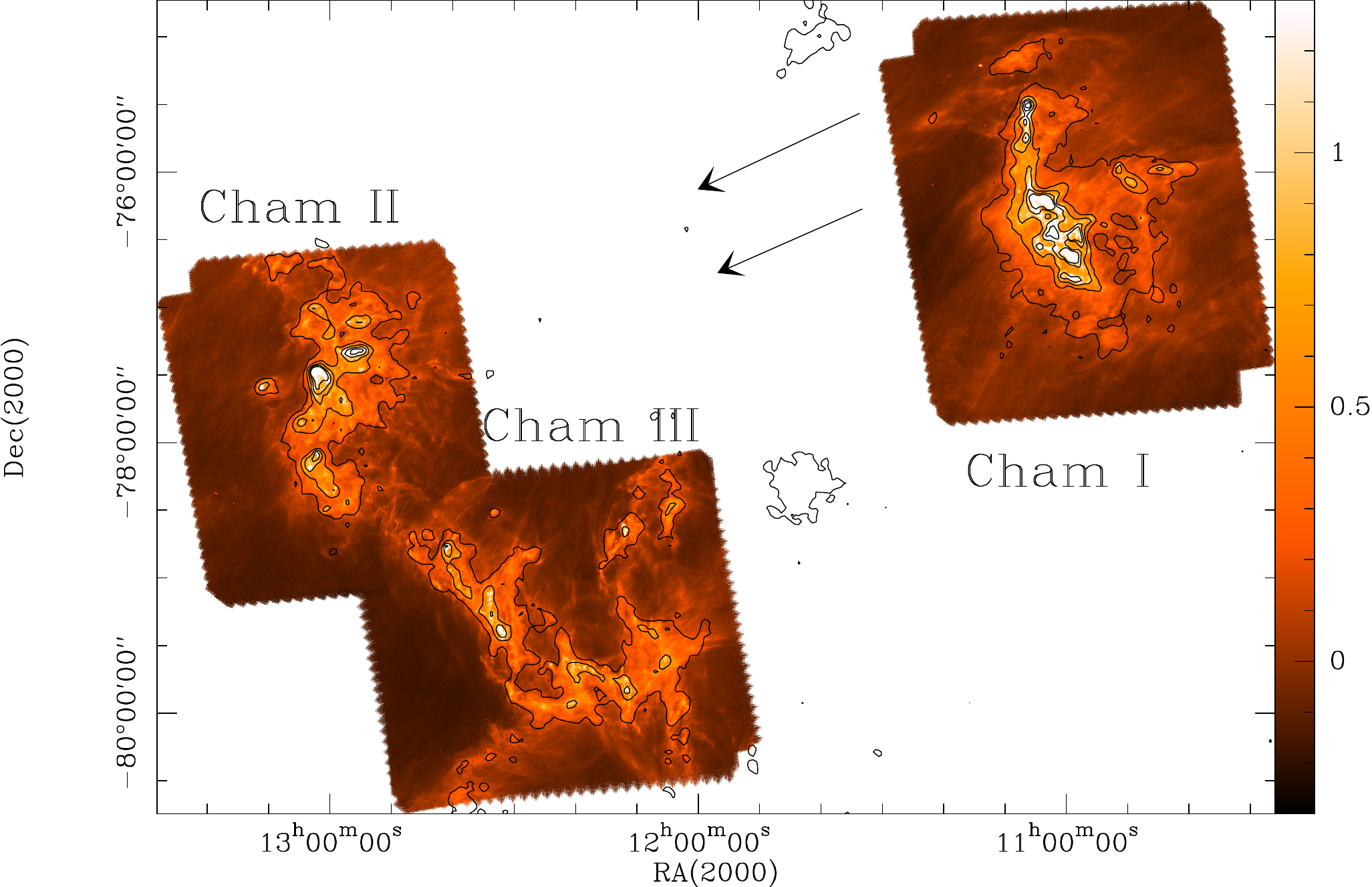}  
\caption[]{SPIRE 250~$\mu$m map with extinction contours \citep{Schneider2011} overlaid from 2 to 10~mag, in steps of 2~mag. The arrows show the general orientation of the magnetic field as measured in Cha~I by \citet{Whittet1994} and \citet{McGregor1994}. Note the preferential alignment of the faint \emph{striations} in Cha~I, less evident in Cha~II or Cha~III where the surrounding diffuse emission takes a variety of orientations. In the region eastwards of Cha~III, no \emph{striations} are detected in the \emph{Herschel} maps. }  
\end{center}  
\label{spire}  
\end{figure*}  
  
Molecular clouds in the Galaxy bear a complex spatial structure highlighted by the unprecedented sensitivity and resolution of the \emph{Herschel} Space Observatory \citep{Pilbratt2010} imaging observations, mainly from large survey programs \citep{Andre2010,Motte2010,Molinari2010}. The recent advances in the census of starless and prestellar cores makes it possible to study their connection to the large scale structure of molecular clouds. Observations show that the prestellar cores identified with \emph{Herschel} are preferentially found within gravitationally unstable filaments \citep[e.g.,][K\"onyves et al., \emph{in prep.}]{Andre2010,Andre2014,Polychroni2013,Konyves2010}, while massive protostellar dense cores and star clusters tend to be found at the junctions of dense filaments \citep[e.g.,][]{Hennemann2012,Schneider2012,Peretto2013}. The observed orientation of filaments suggests that they are aligned with the magnetic field \citep[e.g., in the massive star-forming region DR21,][]{Schneider2010}. Recently, \citet{Palmeirim2013} discovered that low-density filaments (\emph{striations}) in the Taurus region are also preferentially oriented along the magnetic field. These findings suggest that the initial conditions that favour star formation are closely linked to the spatial structure of a molecular cloud. Identifying the processes responsible for the fragmentation of clouds into dense filaments and their subsequent evolution, is therefore paramount to the understanding of core formation, and ultimately, the dependence of core and stellar masses on the large-scale properties of the interstellar medium.  
  
The Chamaeleon molecular cloud complex is one of the nearest star-forming sites located at a distance of 150$-$180~pc \citep{Whittet1997}. It contains the Cha I, II, and III (Fig.~\ref{spire}), as well as the Musca clouds (Cox et al., \emph{in prep.}). \citet{Mizuno2001} showed in their $^{12}$CO~(1$\to$0) survey that the complex is spatially and kinematically coherent with emission in the range of $-$4 to 6~km~s$^{-1}$. Cha~I is the most active star-forming region with a young stellar population of over $\sim$200 members \citep{Luhman2008,Winston2012} with a median age of $\sim$2~Myr. Cha~II ($\sim$4~Myr) has a smaller population of $\sim$60 young stellar objects (YSOs) \citep{Spezzi2008,Spezzi2013}, and no YSO has been found in Cha~III \citep{Belloche2011a}. Using the Large APEX Bolometer Camera (LABOCA), \citet{Belloche2011a,Belloche2011b} mapped the dust continuum emission at 870~$\mu$m in Cha~I and III to single out the possible causes of such different star-formation activity. In Cha~I, the low number of candidate prestellar cores and protostars, as well as the high global star formation efficiency, were interpreted as signs that star formation might be at its end in this cluster, whereas in Cha~III evidence for the on-set of star formation was found.   
  
The Chamaeleon dark clouds were observed with \emph{Herschel} as part of the Gould Belt survey \citep[hereafter, HGBS,][]{Andre2010}. These observations are presented in Sect.~\ref{observations}, and represent an homogeneous dataset in the far-IR across a large area of the cloud complex that allows us to characterize the large-scale structure and extended emission of the three clouds, to better understand their different star formation history. In Sect.~\ref{methods}, we present the different analysis tools employed, such as filament tracing, power spectra, $\Delta$-variance, and probability distribution functions of the column density (PDFs). The properties derived, such as filamentary \emph{vs.} clumpy structure, level of turbulence and energy injection, density structure, are presented in Sect.~\ref{results} and discussed in Sect.~\ref{discussion}, with the conclusions presented in Sect.~\ref{conclusion}.  
  
\begin{figure*}[ht]  
   \centering  
\includegraphics[height=6.cm]{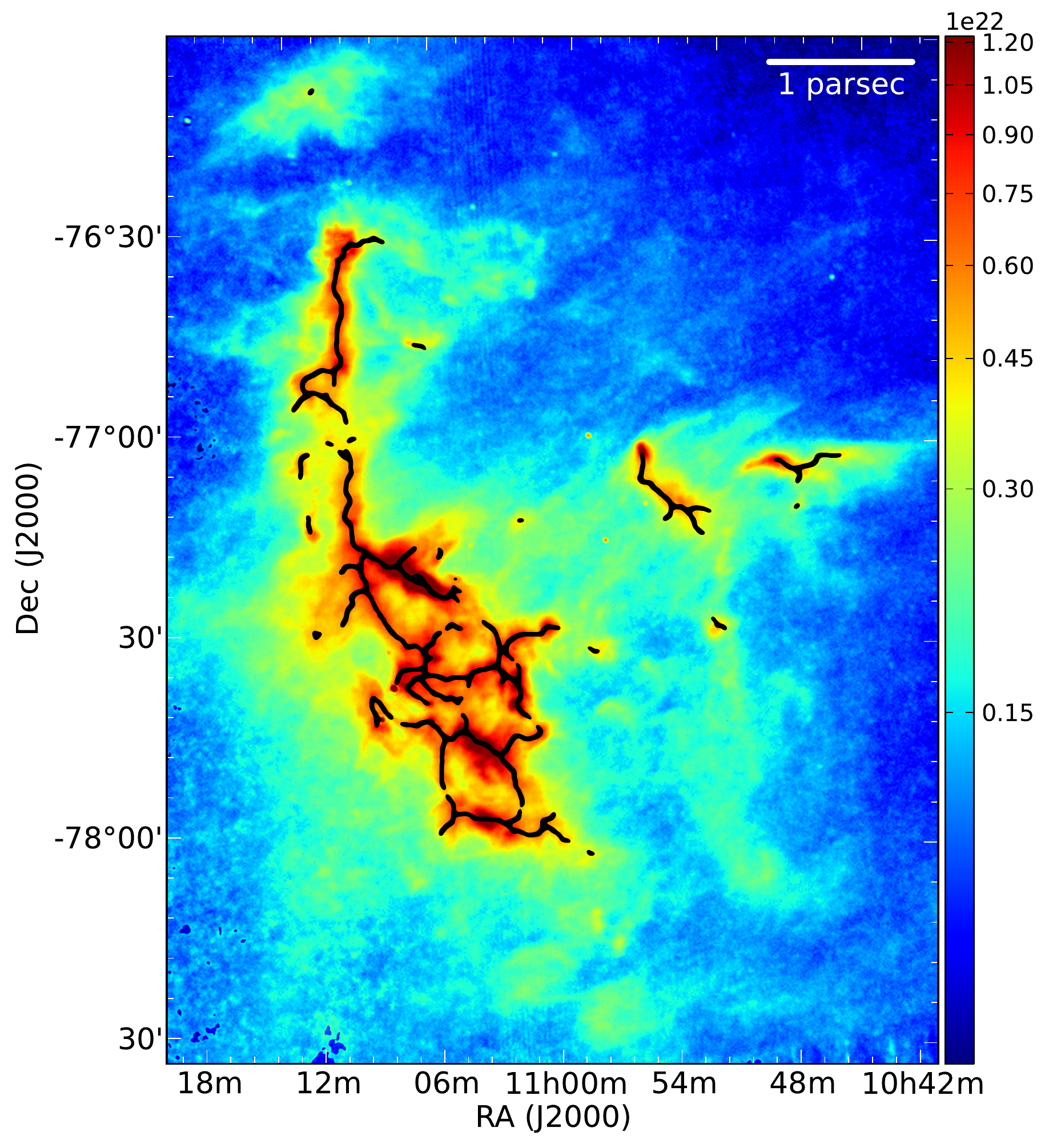}  
\includegraphics[height=6.cm]{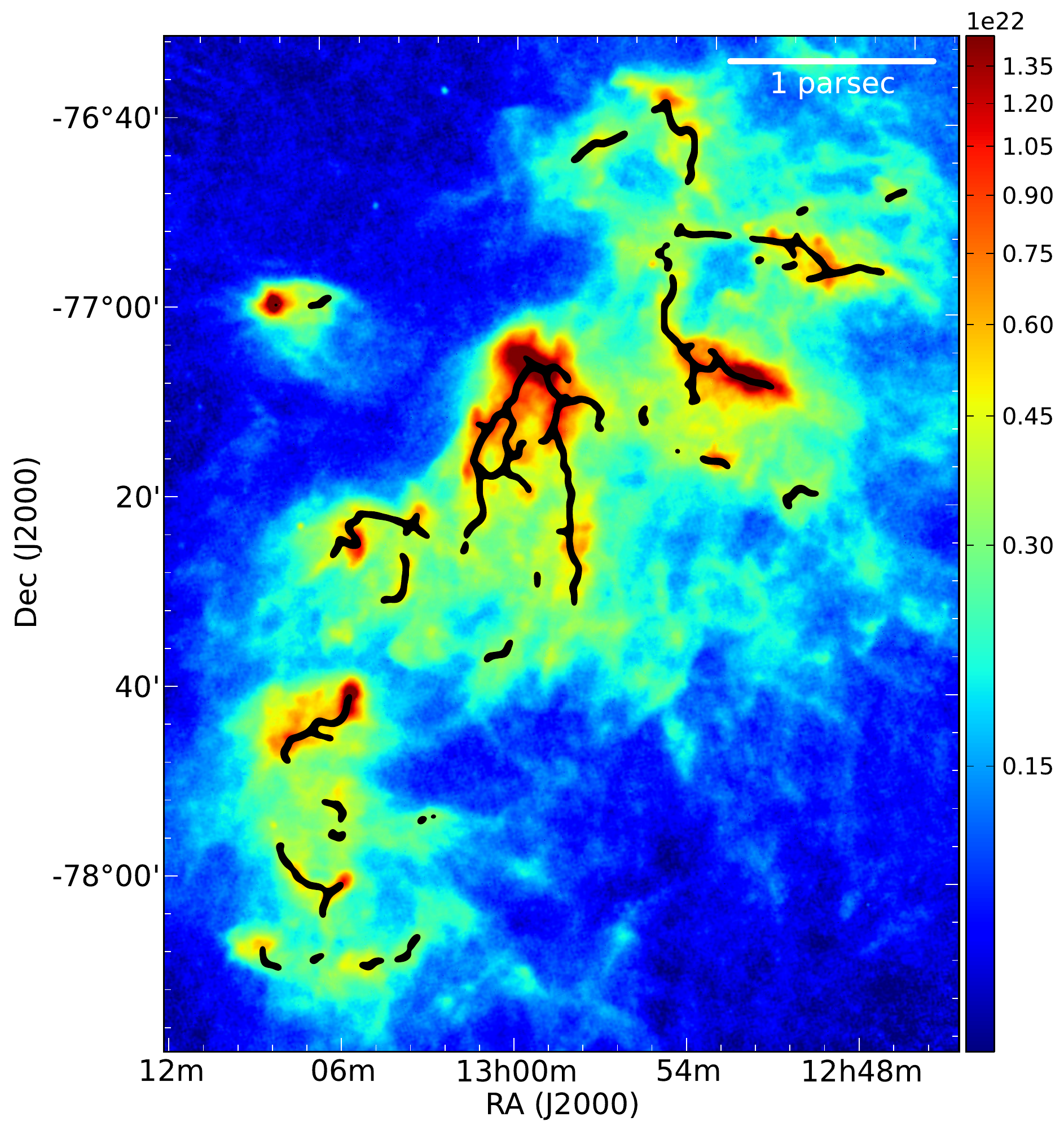}  
\includegraphics[height=6.cm]{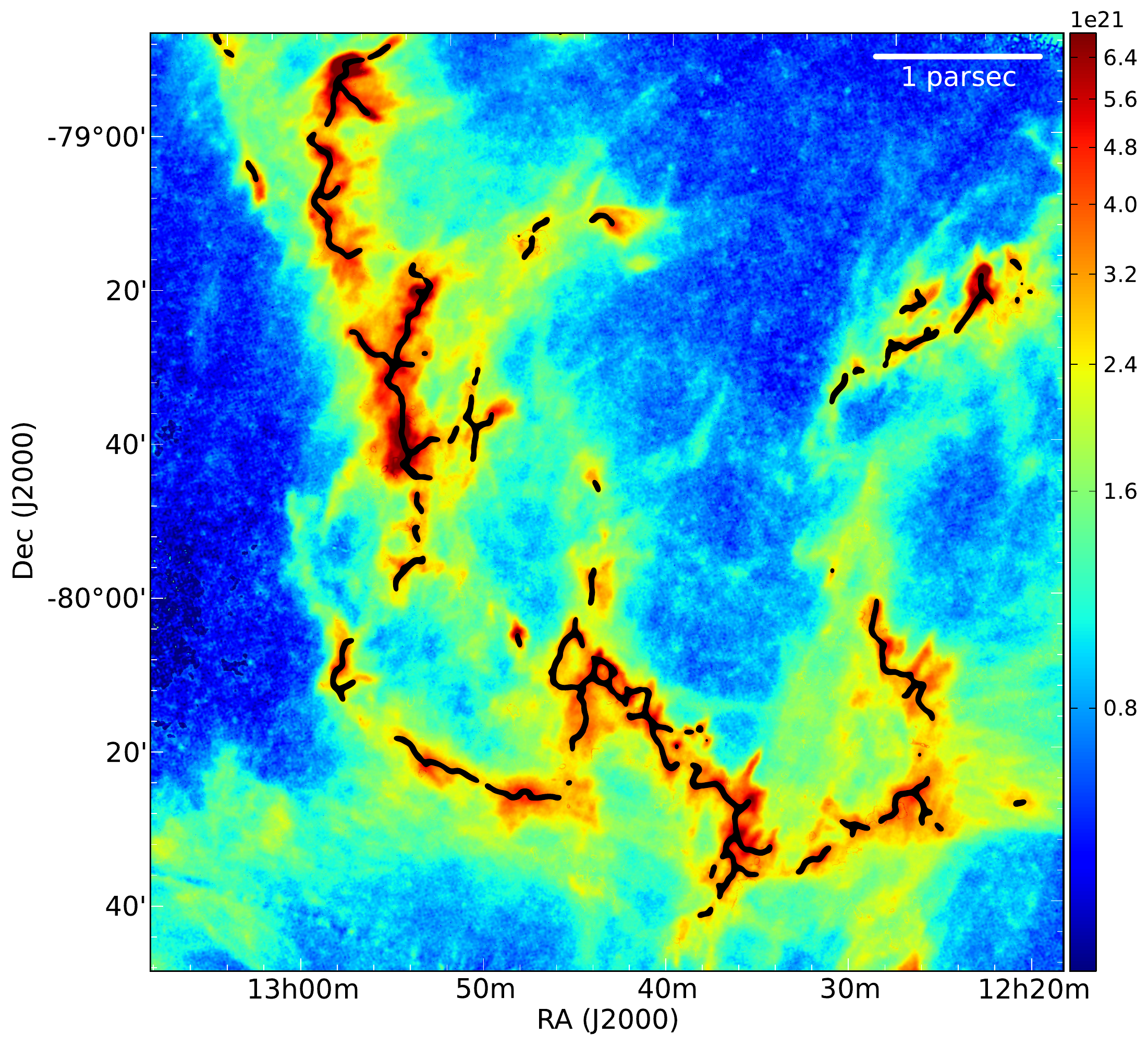}  
\includegraphics[height=6.cm]{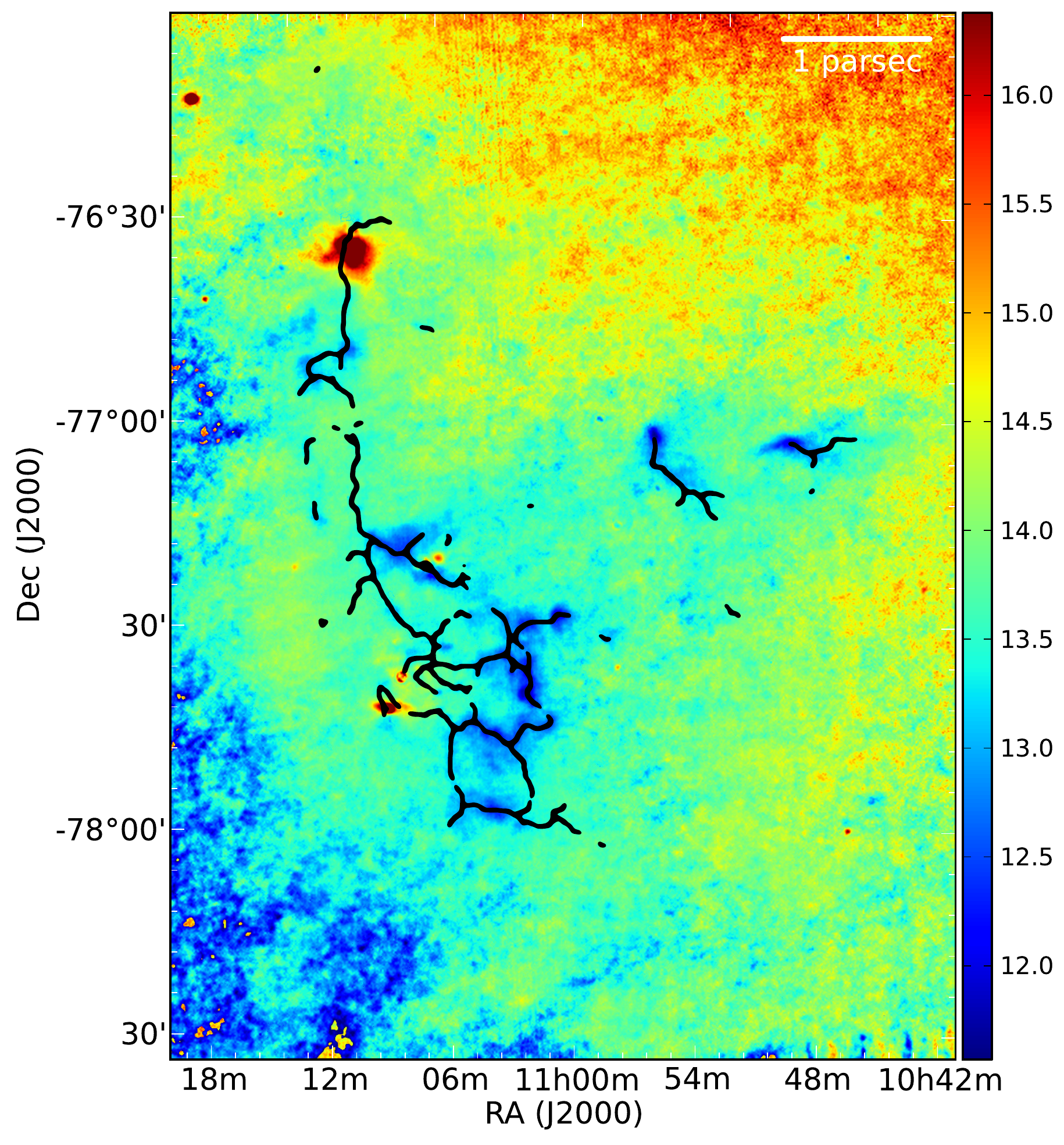}  
\includegraphics[height=6.cm]{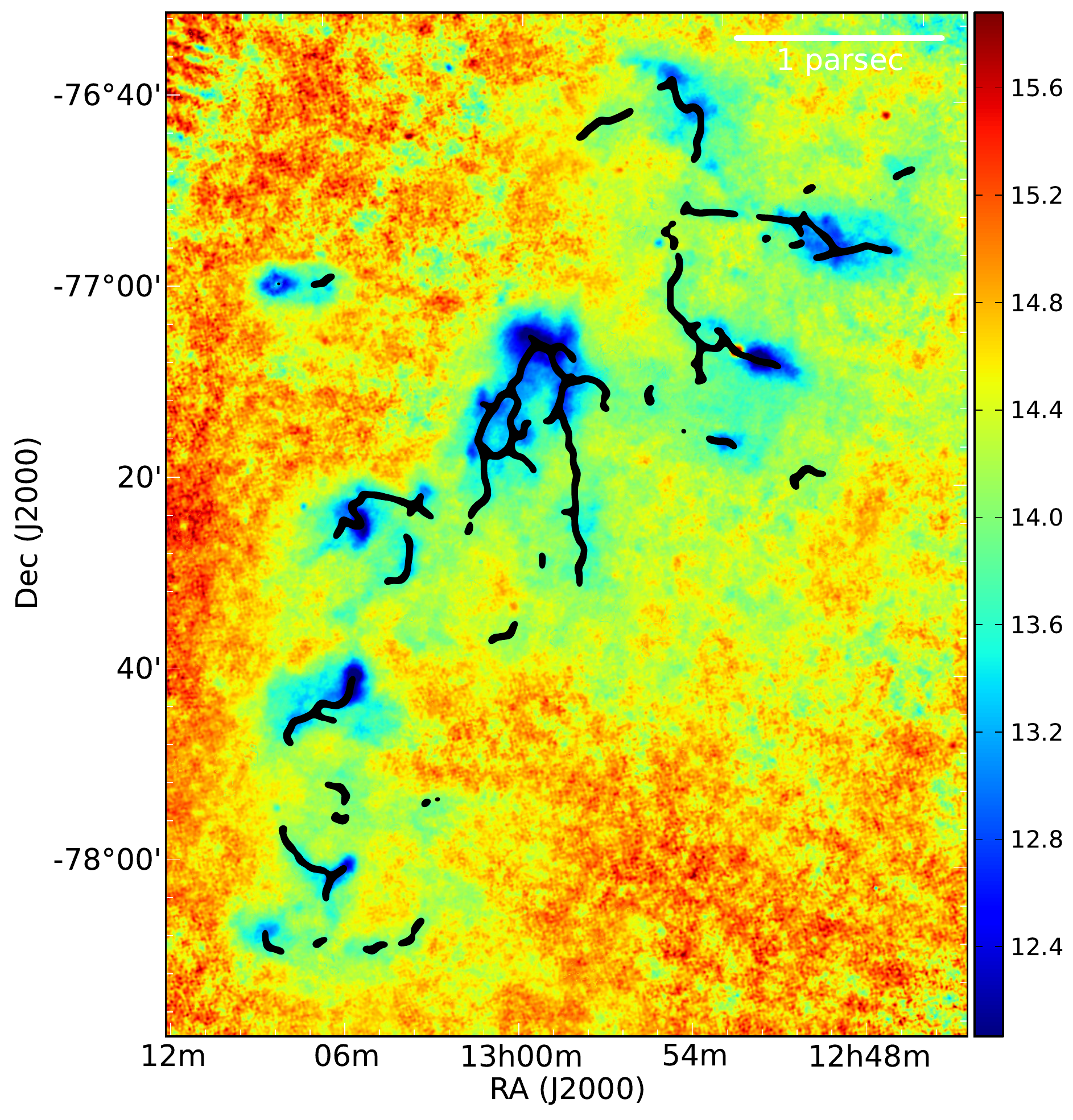}  
\includegraphics[height=6.cm]{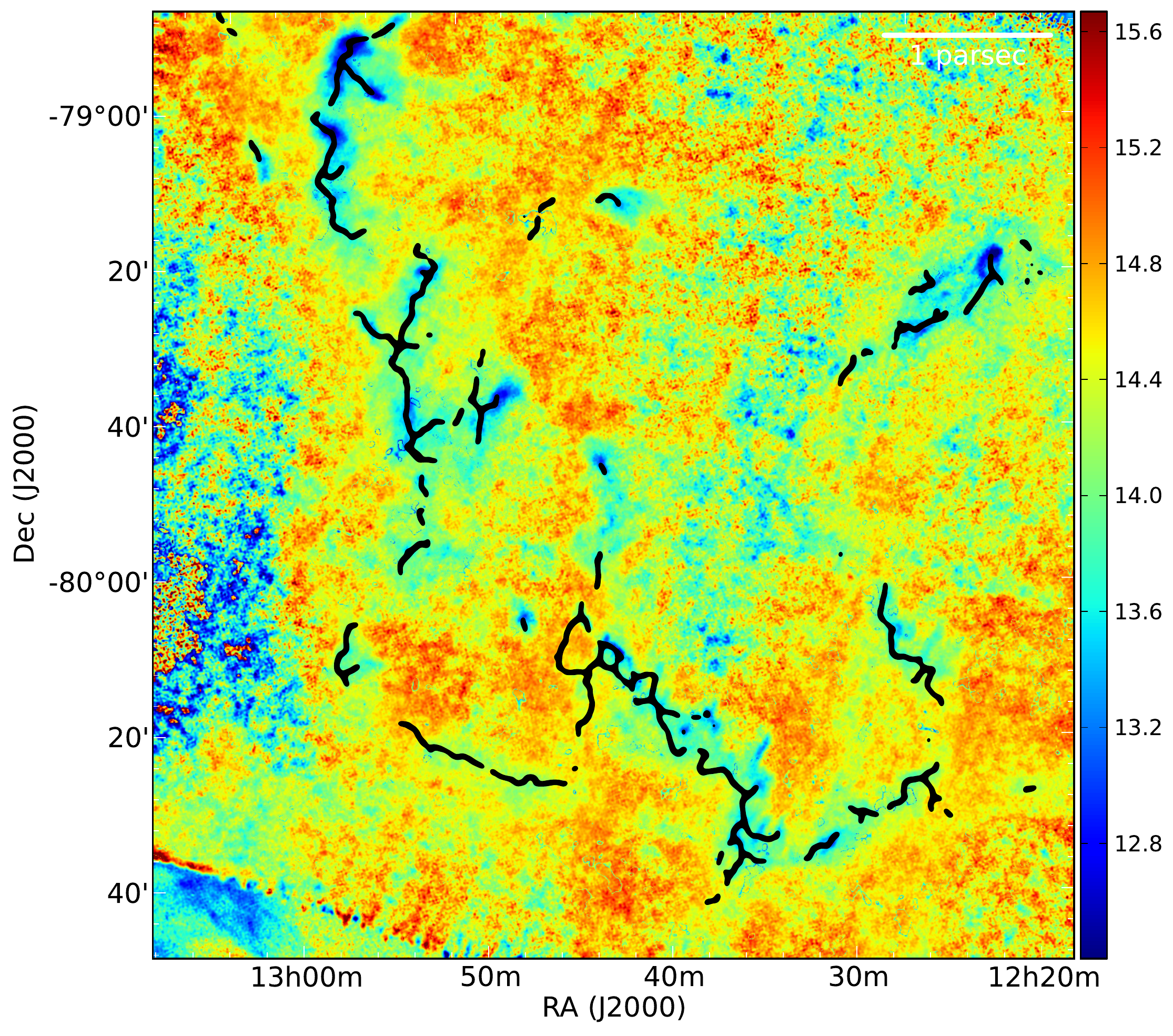}  
\caption{Column density (N$_{H_2}$(cm$^{-2})$) \emph{(top)} and temperature (K) \emph{(bottom)} maps of Cha~I, II, and III \emph{(from left to right)} with the skeletons tracing the most prominent filamentary structure superimposed in black.}  
\label{col}  
\end{figure*}  
  
\section{Observations}  \label{observations}  
The Cha I (obsIDs 1342213178, 1342213179), Cha II (obsIDs 1342213180, 1342213181), and Cha III (obsIDs 1342213208, 1342213209) clouds were observed on the 22 and 23 of January 2011 with the PACS \citep{Poglitsch2010} and SPIRE \citep{Griffin2010} instruments onboard \emph{Herschel} in parallel mode with a scanning speed of 60$''$s$^{-1}$. The angular resolutions for PACS at 160, and SPIRE at 250, 350, and 500~$\mu$m, are $\sim$12$''$, $\sim$18$''$, $\sim$25$''$, and $\sim$37$''$, respectively. The PACS maps at 70~$\mu$m reveal several point sources but do not contain information on extended emission arising from cold dust, and therefore have not been used in this paper.  
  
The reduction of the PACS data was done with {\sl scanamorphos} version 12 \citep{Roussel2013} and is described in \citet{Winston2012} and \citet{Spezzi2013}. The SPIRE data were reduced within the \emph{Herschel} Interactive Processing Environment \citep[HIPE version 7.2,][]{Ott2010}, using a modified version of the pipeline scripts that includes observations taken during the turnaround at the map borders. The two orthogonally-scanned maps were combined using the averaging algorithm `naive-mapper'.  
  
\section{Methods}  \label{methods}  
In the following, we briefly describe the data analysis procedures and tools used to study the \emph{Herschel} observations. Detailed explanations of each method are given in the cited references.  
  
\subsection{Column density and temperature maps} \label{methcol}  
  
The column density and temperature maps were determined from a modified blackbody fit to the 160, 250, 350, and 500~$\mu$m images reprojected to a common 6\arcsec/pixel grid, following the procedure detailed in \citet{Konyves2010}. The zero-offsets, determined from comparison with \emph{Planck} and IRAS data \citep{Bernard2010}, were first applied to each band. For the region covered by PACS and SPIRE simultaneously, we adopted the same opacity law as in earlier HGBS papers \citep{Konyves2010}, fixing the specific dust opacity per unit mass (dust+gas) to be the power-law $\kappa_\nu~=~0.1~(\nu/1000 GHz)^\beta$~cm$^{2}$/g with $\beta$=2 \citep[e.g.,][]{Hildebrand1983}, and leaving the dust temperature and column density as free parameters. The final error on the column density is statistical (SED fitting, photometry) and systematic (opacity law). A recent study by \citet{Roy2014} concluded that the dust opacity law adopted by HGBS is good to better than 50\% accuracy in the whole range of column densities between $\sim3\times10^{21}$ and $10^{23}$~cm$^{-2}$. Figure~\ref{col} shows the resulting maps for the three regions.  
  
\subsection{Filamentary structure} \label{methfil}  
We have used the DisPerSe algorithm \citep{Sousbie2011a,Sousbie2011b} to trace the crest of the filamentary structure of the clouds, using the curvelet component \citep{Starck2003} of the column density map as an input. This method has been successful in mapping the filamentary network in other regions observed with \emph{Herschel}, and the details can be found, for example, in \citet{Arzoumanian2011} or \citet{Schneider2012}. The structure obtained by requiring a persistence threshold of 5$\times$10$^{20}$ cm$^{-2}$ \citep[$\sim$5~$\sigma$, see][for the formal definition of `persistence']{Sousbie2011a} in the curvelet image is overlaid on the column density maps shown in Fig.~\ref{col}. It should be noted that the filamentary structure displayed is derived from the column density map which is a 2D-projection of the volume density. Since DisPerSE works topologically, it connects all emission features such that projection effects may create links between filaments that are not physically related. However, the high spatial resolution of the \emph{Herschel} maps in nearby regions largely alleviates these effects. For example, \citet{Andre2014} (see their Fig.~2) highlight the good agreement between the fine structure of the \emph{Herschel} column density filaments and the C$^{18}$O results of \citet{Hacar2013} for Taurus B211/B213.
 
We characterised the  identified filamentary structures by deriving the radial column density profile perpendicular to the tangential direction to the filament's crest. The centre of each profile was fitted with a \emph{Gaussian}, from which the FWHM and area were taken to determine the filament width and the mass per unit length (Fig.~\ref{distri}).  
  
\subsection{Probability distribution functions of the column density} \label{methpdf}  
  
Probability distribution functions of the (column) density (hereafter 
PDFs) characterise the fraction of gas with a column density $N$ in 
the range [$N$, $N$+$\Delta N$]. For the Chamaeleon clouds, the PDFs 
are represented by the distribution of the number of pixels per 
  log bin versus the column density, expressed in visual 
  extinction (Fig.~\ref{figpdfs}). They are widely used both in 
observational 
\citep[e.g.,][]{Kainulainen2009,Schneider2012,Schneider2013} and 
numerical \citep[e.g.,][and references therein]{Federrath2012} studies 
of the (column) density structure of molecular clouds. For example, \citet{Klessen2000} showed that isothermal, hydrodynamic turbulence 
simulations produce perfectly lognormal PDFs and indeed, probability distribution 
functions obtained from extinction maps \citep{Lombardi2008,Kainulainen2009,Froebrich2010} 
are lognormal for low extinctions. However, \emph{Herschel} observations of the Polaris cloud \citep{Menshchikov2010,Miville2010} show that even in a clearly turbulence dominated cloud, the PDF is not simply lognormal but starts to show excess at higher extinctions \citep{Schneider2013}. Though gravity, in general, plays an important role in organising the  
density structure of a cloud (see below), in a very low-density cloud  
such as Polaris, the excess seen in the PDF can also be due to  
intermittency. Non-isothermal flows can cause a power-law tail  
in the PDF \citep{Passot1998}, but we suspect that  
neither in Polaris nor in the Chamaeleon clouds the temperature  
difference (only a few K) is enough to provoke such flows.  
The most dominant process to influence the PDF is thus gravity, where large-scale collapse as well as individual core 
collapse determine the density structure and give rise to the observed 
power-law tail \citep[see, e.g., numerical simulations 
by][]{Ballesteros-Paredes2011,Kritsuk2013,Girichidis2014}. In 
addition, recent studies \citep[][and Tremblin et al. 2014]{Schneider2012,Rivera-Ingraham2013} showed that radiative 
feedback processes can have a large impact as well, leading to double 
peaks in the PDF and a two-step power law in the tail. The role of the 
magnetic field is, however, not yet clear. MHD simulations show that 
the presence of magnetic fields provokes a more filamentary density 
structure \citep{Hennebelle2013} and a narrower PDF 
\citep[e.g.,][]{Federrath2010}. Observationally,  
a variation on the width of PDFs due to a magnetic field has not yet been observed 
\citep[e.g.,][]{Schneider2013}.
  
PDFs are an important tool to disentangle these various processes. To 
first order, the density PDF is proportional to the column density PDF 
\citep{Brunt2010}. \citet{Ballesteros-Paredes2011} showed that PDFs 
vary during cloud evolution, with purely lognormal shapes found in an 
initially turbulent, non-gravitating cloud, while one or more 
lognormal PDFs at low column-densities and a power-law tail for higher 
values were found for later stages of cloud evolution.  Fitting the 
slope of the high-density tail of the PDF 
\citep{Federrath2011,Girichidis2014} enables the computation of the 
exponent $\alpha$ of the spherical density distribution 
$\rho(r)=\rho_0~(r/r_0)^{-\alpha}$ and thus the relative contribution 
of turbulence and gravitation in a cloud. In the standard 
  inside-out collapse model of a singular isothermal sphere \citep{Shu1987}, $\alpha=2$ is consistent with the collapse of a 
  centrally condensed sphere such as a pre-stellar core, while 
  $\alpha=1.5$ applies to collapsing protostars, i.e. including 
  envelope collapse. In any case, a value of $\alpha$ between 1.5 and 
  2 indicates free-fall collapse of individual cores/protostars and/or 
  and ensemble of cores as was shown in the analytic study of the link 
  between PDF and self-gravity by \citet{Girichidis2014}. However, it is not (yet) analytically possible to 
  differentiate in the slope of the power-law tail between local, 
  small scale core/envelope collapse and larger scale, global collapse 
  of clumps and filaments.  
 
\subsection{Power spectra and $\Delta$-variance}   \label{methpower}
 
To characterise the turbulent structures in the Chamaeleon clouds, we
derive scaling relations such as the power spectrum and the
$\Delta$-variance\footnote{IDL-based routine {\sl deltavarwidget}
  provided by V. Ossenkopf, available at
  www.astro.uni-koeln.de/$\sim$ossk/ftpspace/deltavar.}
\citep{Ossenkopf2008a,Ossenkopf2008b} for each column density map, in
order to measure the relative structural variation as a function of
the size scale. The power spectrum $P(k)$ of a 2-dimensional image
characterizes the injection of energy depending on the wavenumber
(spatial frequency) $k$ with $P(k)~\propto~\vert~k~\vert^{\gamma}$. A
power-law fit to $P(k)$ gives the slope $\gamma$. The $\Delta$-variance
($\sigma_{\Delta}$) is directly linked to the power spectrum
by $\sigma^2_{\Delta}~\propto~L^{\beta~-2}$, where $\gamma~=~1-\beta$
and $L$ is the scale size \citep{Stutzki1998}. Fitting the
slope provides a measure for the amount of structure on various scales
in a given image. A dominance of small scale structure thus implies a
steep slope and high values for $\beta$, while large-scale structures
cause flat $\Delta$-variance curves. The $\Delta$-variance is
  determined purely in the spatial domain and thus limits edge-effect
  problems using Fourier-transform methods, such as power-spectra. It
  effectively separates real structure from observational artefacts since it
  uses the error map as a weight, and it is only limited by the
  spatial resolution and extent of the map. The effect of applying
the $\Delta$-variance to two-dimensional projections of a
three-dimensional structure (in this case, the \emph{Herschel} column
density maps) has been addressed by \citet{MacLow2000}, who
demonstrated that there is a simple translation between a projected
$\Delta$-variance and a three-dimensional
$\Delta$-variance.

\begin{figure*}  
   \centering  
\hspace{0.0cm}\includegraphics[width=0.9\columnwidth]{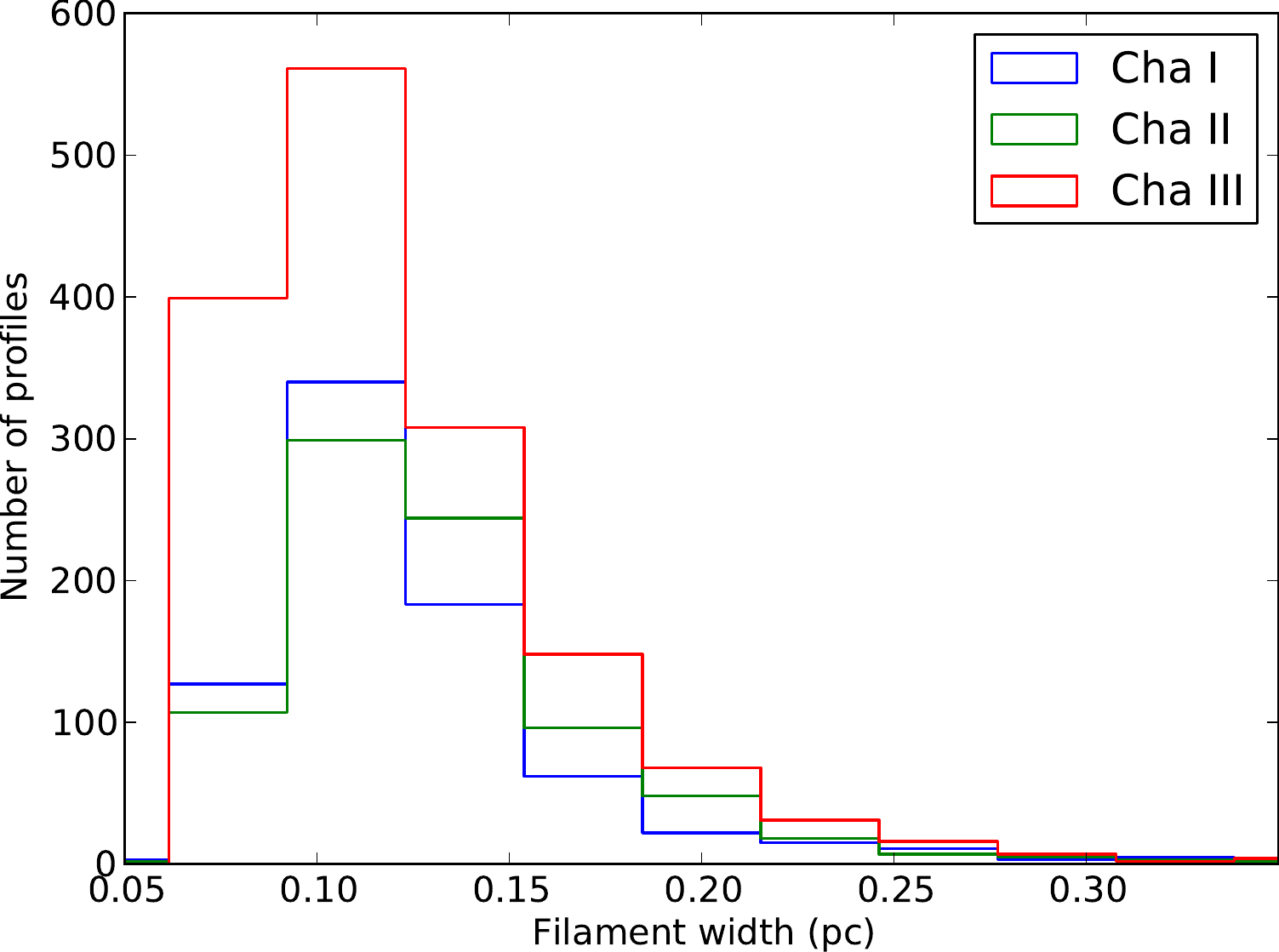}  
\hspace{1.0cm}\includegraphics[width=0.9\columnwidth]{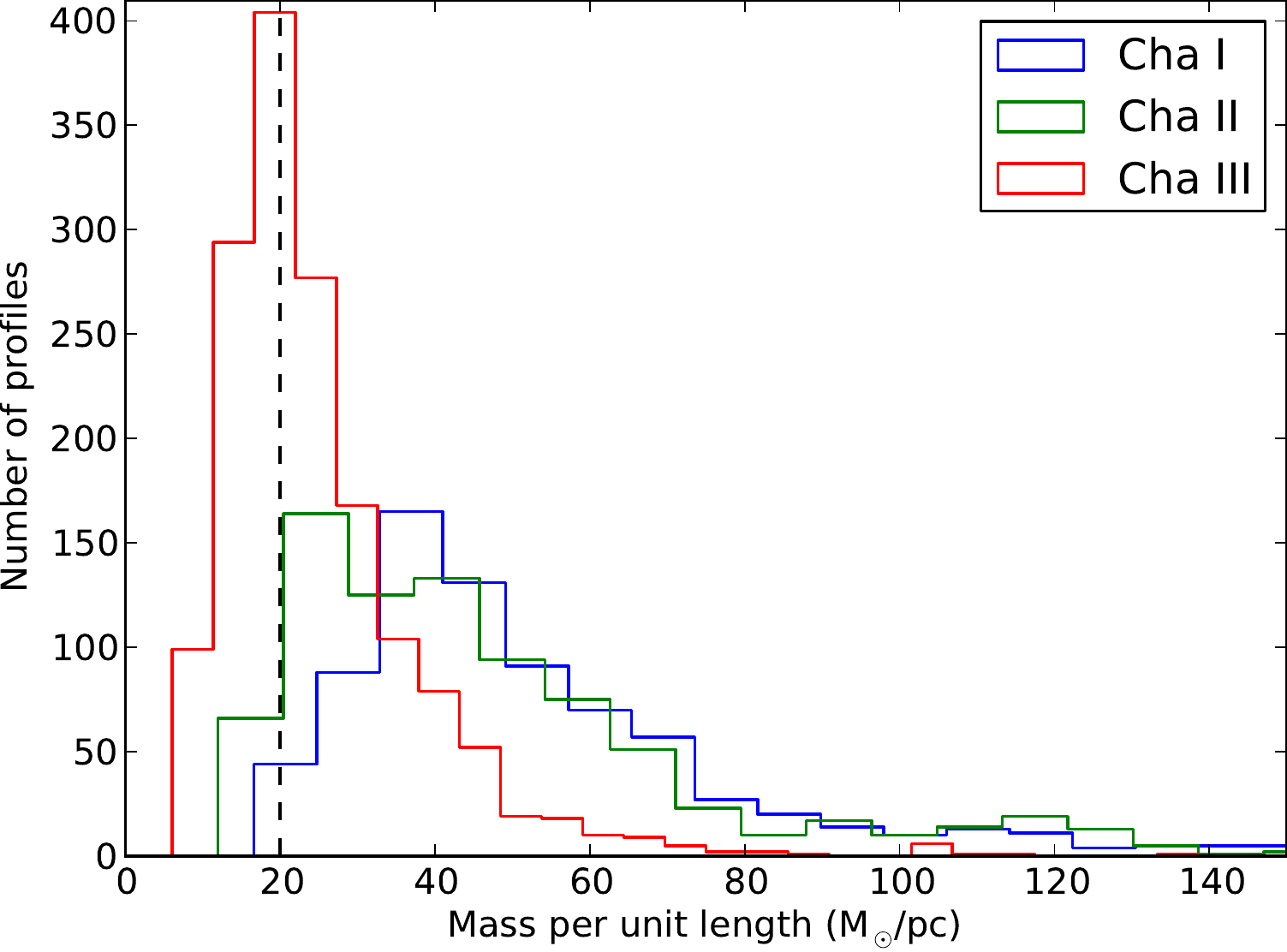}  
 \caption{Distribution of the width \emph{(left)} and mass per unit length \emph{(right)} for individual profiles, as derived for Cha~I (blue), Cha~II (green), and Cha~III (red). On the mass per unit length histogram, the dashed line shows the thermal critical mass per unit length for a gas temperature of $\sim$12~K.}  
\label{distri}  
\end{figure*}  
  
\section{Results} \label{results}  
  
\subsection{Column density and temperature maps} \label{colum}  
   
A comparison between the \emph{Herschel} column density map (Fig.~\ref{col}) and the near-IR extinction map \citep{Schneider2011} shows both to be consistent in structure as well as intensity (the maximum column density is $\sim$7$\times$10$^{21}$ cm$^{-2}$ for Cha~I, see Table~\ref{table-col}). The detailed spatial structure is  
different for the three clouds. Cha~I is dominated by a central \emph{ridge} of emission, i.e., an elongated cloud structure of relative higher density, surrounded by faint \emph{striations}, reminiscent of those found in the B211/3 filament in Taurus by \citet{Palmeirim2013}. These cold (11--13 K) features are also clearly seen in the SPIRE 250~$\mu$m map (Fig.~\ref{spire}). A comparison with C$^{18}$O~1$\to$0 data \citep[see Fig.~2 in][]{Haikala2005} shows that the structures seen in the column density map consist of overlapping filaments, that are coherent in  
velocity, and individual clumps. The only hot spot in all regions is the dust ring around HD~97300 located in Cha~I, with T$\sim$25~K \citep{Kospal2012}. Cha~II has a more fragmented appearance (best seen in Fig.~\ref{col}) with extended emission regions. In the temperature maps, individual \emph{clumps} are easily identified as patchy, cold (11--14 K) regions. The most clearly defined \emph{filamentary} region (large aspect ratio of the identified structures) is Cha~III. Here, as well as in the other regions, the cold denser regions are embedded in a lower density (1--3$\times$10$^{21}$ cm$^{-2}$) background.  
  
The total masses derived from the column density maps are given in Table~\ref{table-col}. The values are lower than the ones estimated from $^{12}$CO~(1$\to$0) observations \citep{Mizuno2001} by factors of 1.2, 3.8, and 7.8 for Cha~I, II, and III, respectively, and we attribute the discrepancy mainly to the unreliable $^{12}CO/H_{2}$ conversion factor. The \emph{Herschel}-derived mass for Cha~I of 865~M$_\odot$ is higher than the one calculated from C$^{18}$O~(1$\to$0) data \citep[230~M$_\odot$,][]{Haikala2005}. This difference has been noted by \citet{Belloche2011a}, where it is shown that C$^{18}$O~(1$\to$0) is not a good gas tracer below an A$_{V}$~=~6~mag. It is nevertheless a reliable tracer of mass between A$_{V}$~=~6--15~mag in Cha I, when compared to the mass derived from extinction maps \citep{Belloche2011a}. Furthermore, the region surveyed in \citet{Haikala2005} does not include the western part of Cha~I. Indeed, our mass estimate is in agreement with the value derived from the near-IR extinction map \citep{Schneider2011}.  
  
\begin{table}[]   
\caption{Column density and mass values.}  
\begin{tabular}{lccccc}  
\hline  
              & N$_{noise}$ & N$_{max}$ & $<$N$>$ & Mass \\  
Cloud         & (1)         &  (2)      & (3)     & (4)  \\  
\hline  
Cha I    & 0.30        & 7.26      & 1.83    & 865 \\  
Cha II   & 0.31        & 4.35      & 1.59    & 502 \\  
Cha III  & 0.30        & 1.86      & 1.25    & 241 \\  
\hline  
\end{tabular}\\  
\tablefoot{(1,2,3) rms noise, peak, average column density in 10$^{21}$ cm$^{-2}$. (4) Total mass of cloud in M$_\odot$ above a threshold of $A_{V}$=2~mag \citep[$\approx$2$\times$10$^{21}$ cm$^{-2}$, using the conversion formula N(H$_2$)/\av=0.94$\times$10$^{21}$ cm$^{-2}$ mag$^{-1}$][]{Bohlin1978}. The distance of 180~pc was taken for Cha~I and II, and 150~pc for Cha~III.}     
\label{table-col}  
\end{table}

\begin{figure*}[ht]  
\hspace{0.0cm}\includegraphics[angle=90,width=5.5cm,trim=3cm 7cm 2cm 2cm,clip=true]{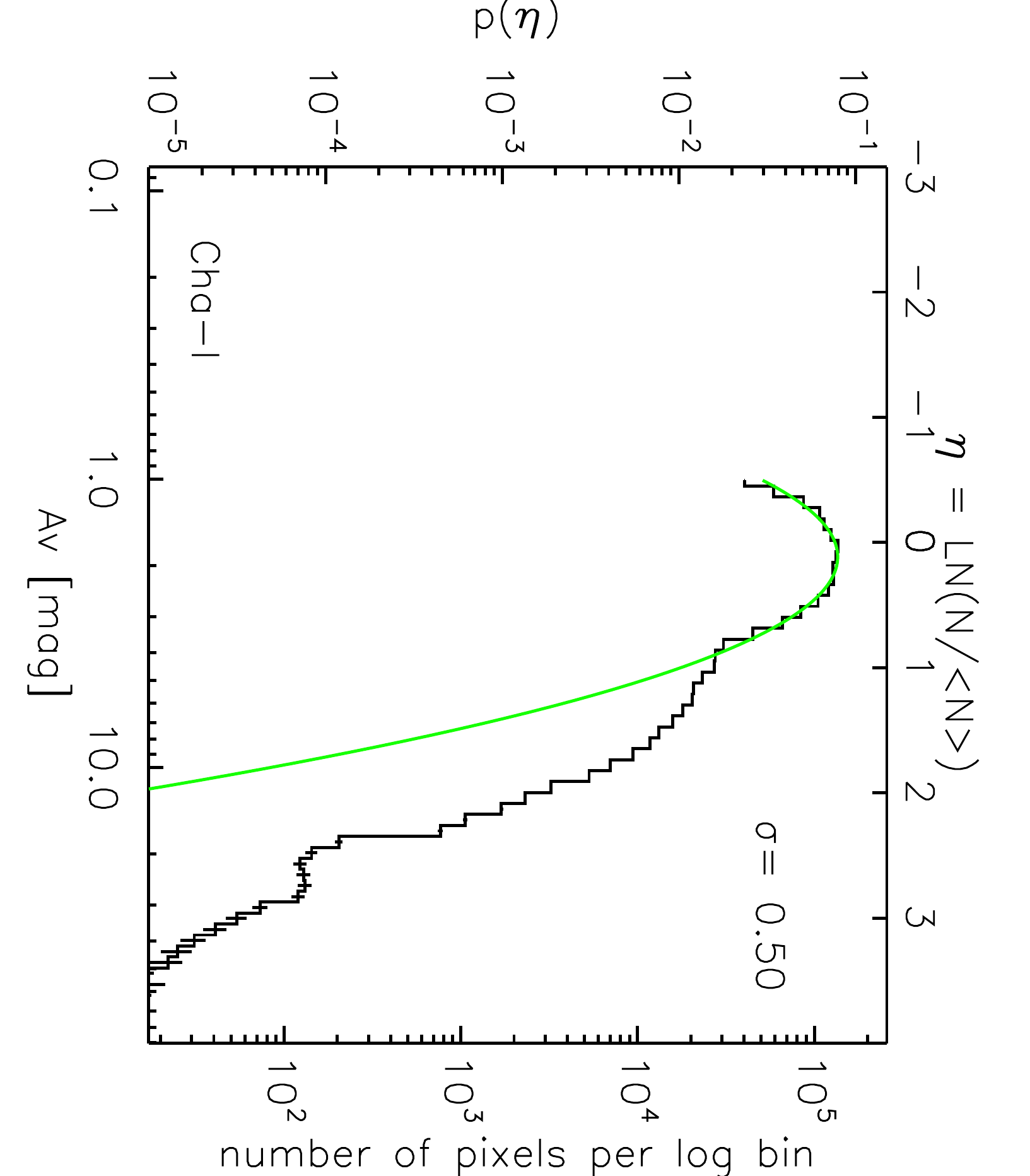}  
\hspace{0.5cm}\includegraphics[angle=90,width=5.5cm,trim=3cm 7cm 2cm 2cm,clip=true]{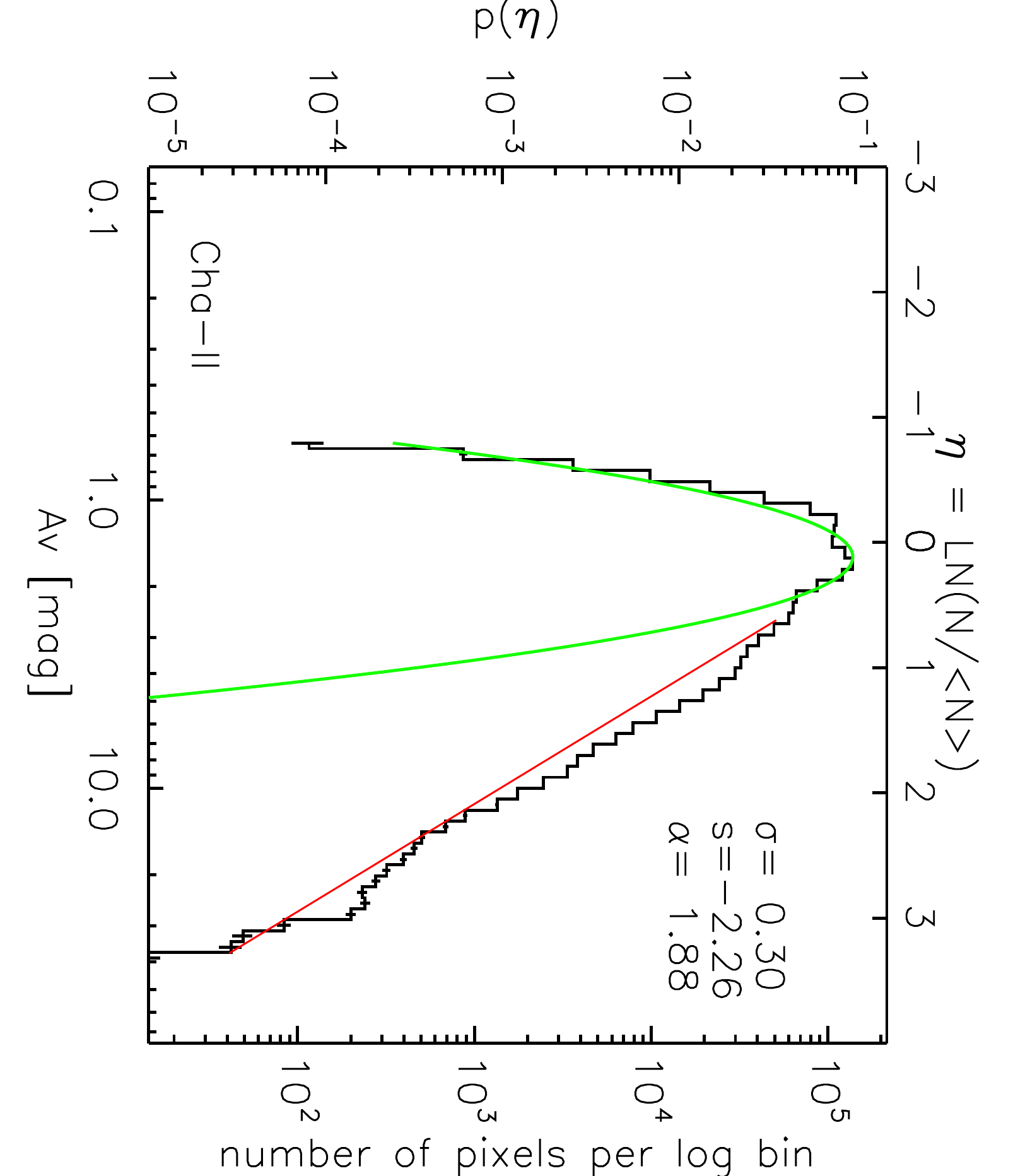}
\hspace{0.5cm}\includegraphics[angle=90,width=5.5cm,trim=3cm 7cm 2cm 2cm,clip=true]{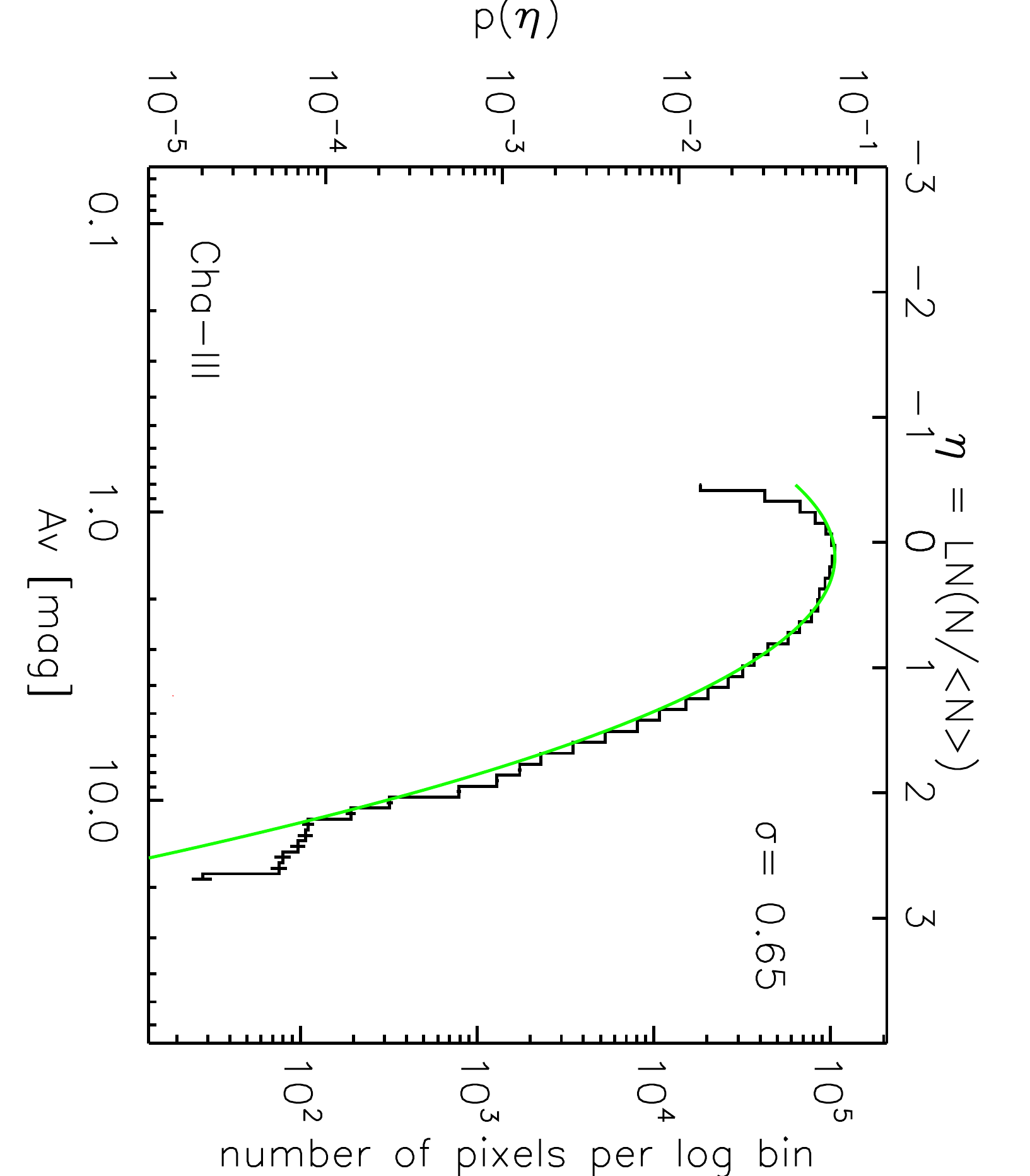}  
\caption{Probability distribution functions of column density for the Cha I, II, and III cloud regions. The PDFs were obtained from the column density maps (Fig.~\ref{col}) at an angular resolution of 36$''$. The left y-axis indicates the normalized probability and the right y-axis the number of pixels per log bin. The green line indicates the fitted PDF and the red line (in Cha~II) the power-law fit to the high-density tail. The width of the PDF ($\sigma$), the fitted slope $s$, and the corresponding exponent $\alpha$ of a spherical density distribution are given in each panel.}  
\label{figpdfs}  
\end{figure*}  
  
\subsection{Filament parameters} \label{resfil}   
The distributions of widths for the profiles of the structure traced by the DisPerSe algorithm where the contrast (ratio of the column density at the crest to local background) was $>$3 are shown in Fig.~\ref{distri} \emph{(left)} for the three clouds. This threshold ensures a sufficient number of profiles, while minimizing contamination from profiles in confused regions. The obtained distributions have a median value 0.115$^{+0.038}_{-0.023}$, 0.124$^{+0.044}_{-0.029}$, and 0.112$^{+0.047}_{-0.026}$~pc (the upper and lower bounds mark the 68\% confidence interval), for Cha~I, II, and III, respectively. This result shows that there is no significant difference in width between the clouds, and is in agreement with the characteristic width measured in other star-formation regions \citep[e.g.,][]{Arzoumanian2011}.  
  
In Fig.~\ref{distri} \emph{(right)}, we show the distribution of the projected mass per unit length for the same profiles. If we take the gas temperature in the Chamaeleon complex filamentary-like structures to be approximately 12~K, the thermal critical mass per unit length is $\sim$20~$M_\mathrm{\sun}/pc$. Gravitationally supercritical filaments are defined as having a mass per unit length greater than this critical value. This implies that most of the structures in Cha~I and II are supercritical and likely to be undergoing collapse, while in Cha~III the majority of the individual profiles where the filaments were probed are found to be subcritical.   
  
\subsection{Probability distribution functions of column density} \label{pdf_cham}   
The PDFs of the column density obtained for the three clouds are displayed in Fig.~\ref{figpdfs}. Common to the three regions is a lognormal distribution for low extinctions with a width of 0.48 to 0.6~mag and a peak at $A_{V}$$\sim$2~mag. For higher column densities, the PDFs show significant differences.  
  
The PDF for {\bf Cha~I} shows a turnover from a lognormal low-density component into a higher density feature around $A_{V}$=4--5~mag. The pixel distribution above these $A_{V}$ values is neither a power-law tail nor a second lognormal PDF (fitting this component with a lognormal distribution is not possible though it may appear as such by eye-inspection).  This scenario resembles that observed in the intermediate/high-mass star-forming region Vela C (Hill et al. 2011) where the same characteristics in the column density map and PDF were attributed to the contrast between a dense and massive ridge, embedded in a lower-density gas component. It is thus possible that the segregation between ridge/bulk emission of a cloud is a common feature in molecular clouds. At even higher column densities (A$_{V}$~$>$~20~mag) there is a small but significant pixel statistics in the PDF that resemble a power-law. Spatially, these pixels comprise the star-forming clumps (see also Sect. 5.2).  
  
The PDF of {\bf Cha~II} shows the clearest example of a power-law tail (though there is some excess in the distribution), starting at A$_{V}$$\sim$3~mag. Fitting a single power-law leads to a slope of $s$~=~--2.26. Assuming that the power-law tail is due to purely spherical collapse, the density distribution varies as $\rho(r)~\propto~r^{-\alpha}$, determined from $s$ with $\alpha=-2/s + 1$ \citep{Federrath2013,Schneider2013}, and the exponent $\alpha$ is estimated to be 1.88, which implies the dominance of self-gravity.
  
The PDF of {\bf Cha~III} can be best fit by a single lognormal 
  across the whole density range though the low column density range (left 
  of the peak) is not perfectly covered.  Some excess at higher 
  column-densities (A$_{V}$~$>$~10~mag) is observed and this feature 
is now more common in `quiescent' molecular clouds, e.g., in Polaris 
\citep{Schneider2013}, or the `Spider' molecular cloud (Schneider et 
al., \emph{in prep.}), pointing towards a scenario where these clouds 
are not only dominated by turbulence (which would lead to a perfectly 
lognormal PDF). For Polaris, it was argued that statistical density 
fluctuations and intermittency may provoke this excess. In Cha~III, 
however, there are first signs of core formation in some filaments so 
that we speculate that gravity could be causing the start of a 
power-law tail distribution. 
  
\subsection{Power-spectra and $\Delta$-variance} \label{turbulence}   
  
\begin{figure*}[ht] 
\centering
\hspace{0.0cm}\includegraphics[angle=0,width=7.6cm,height=9.1cm]{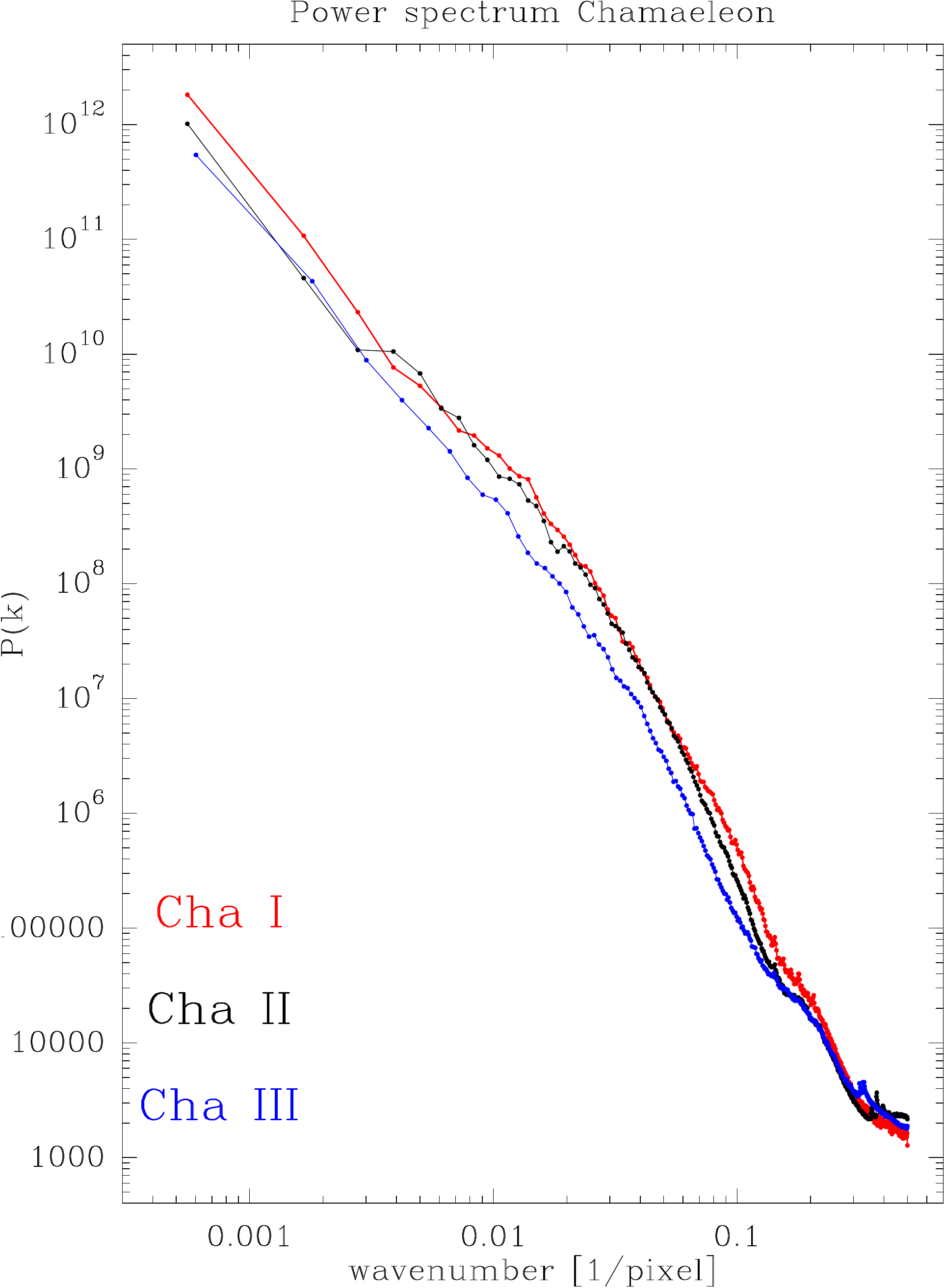}  
\hspace{0.0cm}\includegraphics[angle=0,width=7.6cm,height=9.1cm]{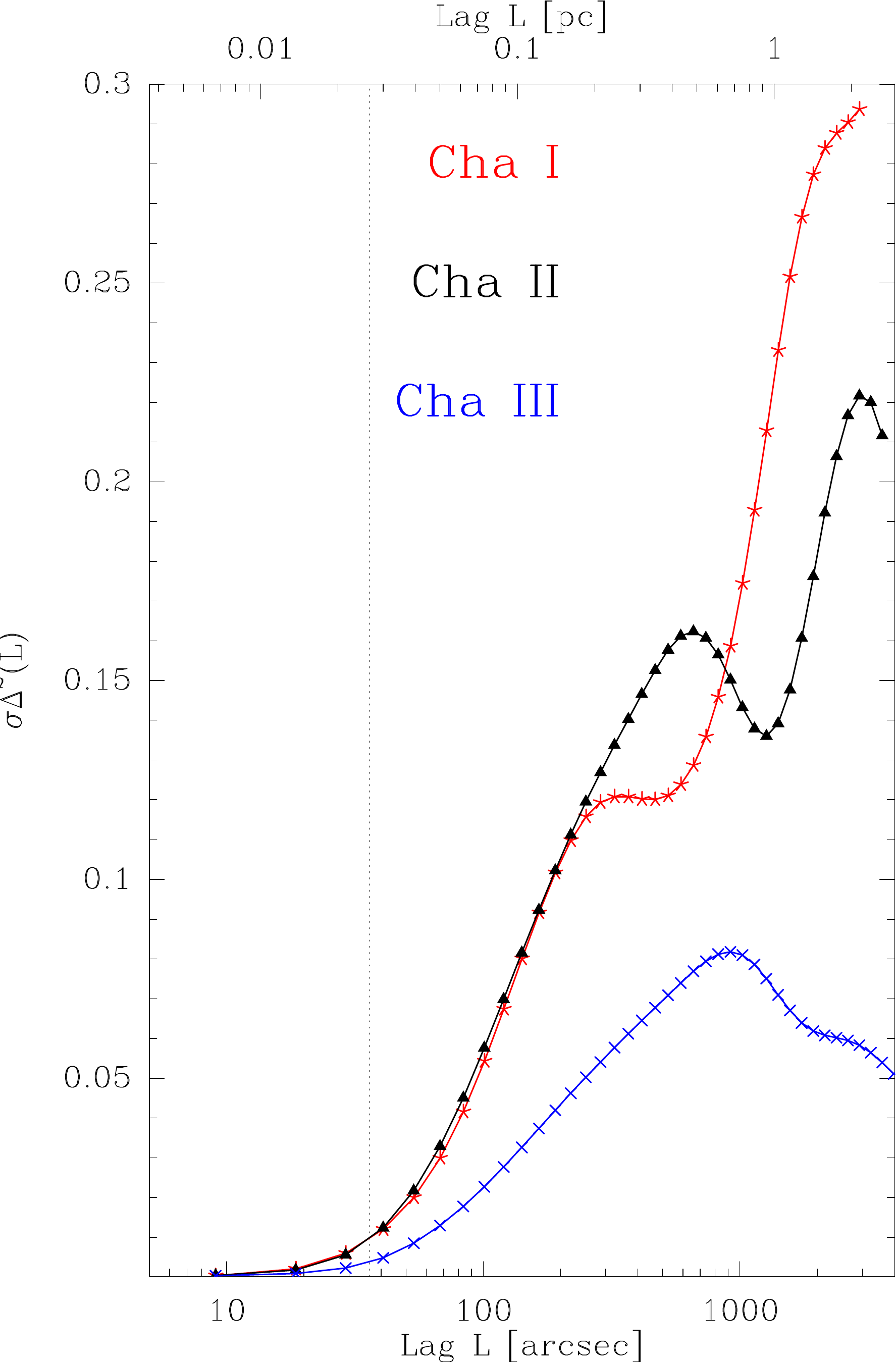}  
\caption[]{(\emph{Left}) Power spectra of the column density maps of the Chamaeleon clouds, which were determined from the \emph{Herschel} images reprojected to a common 6\arcsec/pixel grid (see Sect.~\ref{methcol}). (\emph{Right}) The $\Delta$-variance spectra of the same column density maps. The dotted vertical line indicates the resolution limit of 36\arcsec.}
\label{power}  
\end{figure*}  
  
Figure~\ref{power} shows the power spectra and $\Delta$-variance for the three clouds. As mentioned in Sect.~\ref{methpdf}, the modulations in the power-spectra are generally less obvious than the ones in the $\Delta$-variance so we focus the discussion on the latest.  
  
The $\Delta$-variance of {\bf Cha~I} shows a clear change in slope at 0.15~pc, stays flat until $\sim$0.6~pc and rises again up to 2~pc. We interpret this curve as caused by {\sl cores and clumps} (with a size distribution between 0.2 and 0.5~pc, i.e., not determined by turbulence), and by the {\sl ridge} that gives origin to the second peak. The density distribution seen in the PDFs (Sect.~\ref{pdf_cham}) is thus also traced in the spatial structure.   
  
In {\bf Cha~II}, the $\Delta$-variance appears to be turbulence dominated (no range of a flat $\Delta$-variance spectrum as for Cha~I) but displays a peak at $\sim$0.5 pc likely caused by the typical size scales of the {\sl clumps}, since much less {\sl cores} are detected in Cha~II, embedded in the larger extended ridge-like structure on a size scale of $\sim$2~pc (second peak).  
  
Finally, {\bf Cha~III} shows a $\Delta$-variance with one peak at $\sim$0.7~pc, probably indicating the typical lengths of the filaments, but appears otherwise featureless, consistent with the lack of prominent structures other than the filamentary network.  
  
We note that there is no indication for a typical size scale of 0.1~pc
that would represent the filament width. Instead, the constant rise of
the curves until the first peak in the different regions indicates
that structures on many size scales are present. We assume that in
particular the \emph{striations} with their large variation in size
prevent a clear identification of filament width for the
$\Delta$-variance.  The fitted slope of the $\Delta$-variance up to
0.15~pc (lowest common scale until a turnover occurs), gives
values of $\beta$=2.49$\pm$0.59 for Cha~I,
  $\beta$=2.50$\pm$0.21 for Cha~II, and $\beta$=2.36$\pm$0.23 for
  Cha~III. These values are at the lower end of those found for a
  large compilation of different molecular clouds by
  \citet{Bensch2001} and \citet{Schneider2011}.  The low values
  indicate that the power is contained in the largest scales,
  suggesting that large-scale physical processes govern the structure
  formation in the Chamaeleon complex. These could be an overall
magnetic field or energy injection from supernova explosions.
    
\section{Discussion} \label{discussion}  
\subsection{Comparison to numerical simulations}  
  
In order to compare our results with turbulence models, we computed independently the hydrodynamic Mach-number ($\mathcal{M}$) using  
molecular line data \citep[see][for details]{Schneider2013,Pineda2008}. In contrast to \citet{Belloche2011a}, where C$^{18}$O~1$\to$0 observations were used, we opt for the $^{12}$CO~(1$\to$0) data from \citet{Mizuno2001} with the molecular line FWHM between 2 and 2.8 km~s$^{-1}$, since only this line traces the low-density gas component and thus the bulk of the molecular cloud. In the calculation we assume LTE and, additionally, that the gas and dust are thermally coupled, taking the maximum brightness temperature at the peak of the line to be the dust temperature derived from our \emph{Herschel} maps ($\sim$14~K). Given these assumptions, we caution that the Mach number calculation carries a large uncertainty. We derived values of $\mathcal{M}$=7.2, 10.0, and 9.3 for Cha~I, II, and III, respectively.
 
A comparison between the observed PDFs and hydrodynamic simulations 
 including gravity, different turbulent states 
with a range of Mach-number between 2--50, and star-formation 
efficiencies (SFE) from 0 to 20\% \citep{Federrath2012}, results in fits to models where 
$\mathcal{M}$=3, 5, or 10 with solenoidal or mixed forcing (with or 
without magnetic field) and low SFE. Furthermore, the case of Cha~III 
(mostly lognormal, showing only a slight deviation at high densities, 
though not a clear power-tail) is only reproduced by a SFE of 0\% and 
different values of the Mach-number (the higher $\mathcal{M}$ the 
broader the PDF). The Cha~II region cannot be clearly pinpointed. 
Various combinations of $\mathcal{M}$ and SFE lead to this PDF shape. 
However, they share the fact that the SFE is higher than 0\% and that 
compressive forcing is not the dominating source of energy because the 
PDF is narrow. Here, we make use of the results of the 
$\Delta$-variance that led to slope values of $\beta$~=~2.3--2.5 (with 
an error of 0.2 to 0.6) for all regions, corresponding to an 
$\gamma$~=~--1.3 to --1.5 for the exponent of the radial 
  density distribution. Comparing these values to Table~2 in 
\citet{Federrath2013} shows that the SFE for these values lies between 
0 and 1\%. The value of $\alpha$~=~0 given in \citet{Federrath2013}, 
that was determined from the extinction map presented in 
\citet{Schneider2011}, is consistent with our values, considering our 
large error on $\beta$. The case of Cha~I is only vaguely reproduced 
in some models (e.g., solenoidal forcing with SFE 5\% and 
$\mathcal{M}$=10) where a cut-off at high densities due to resolution 
effects may also play a role. 
  
\subsection{The peculiar density distribution in Cha~I}   
\begin{figure}  
   \centering  
\hspace{0.0cm}\includegraphics[angle=0,width=\columnwidth]{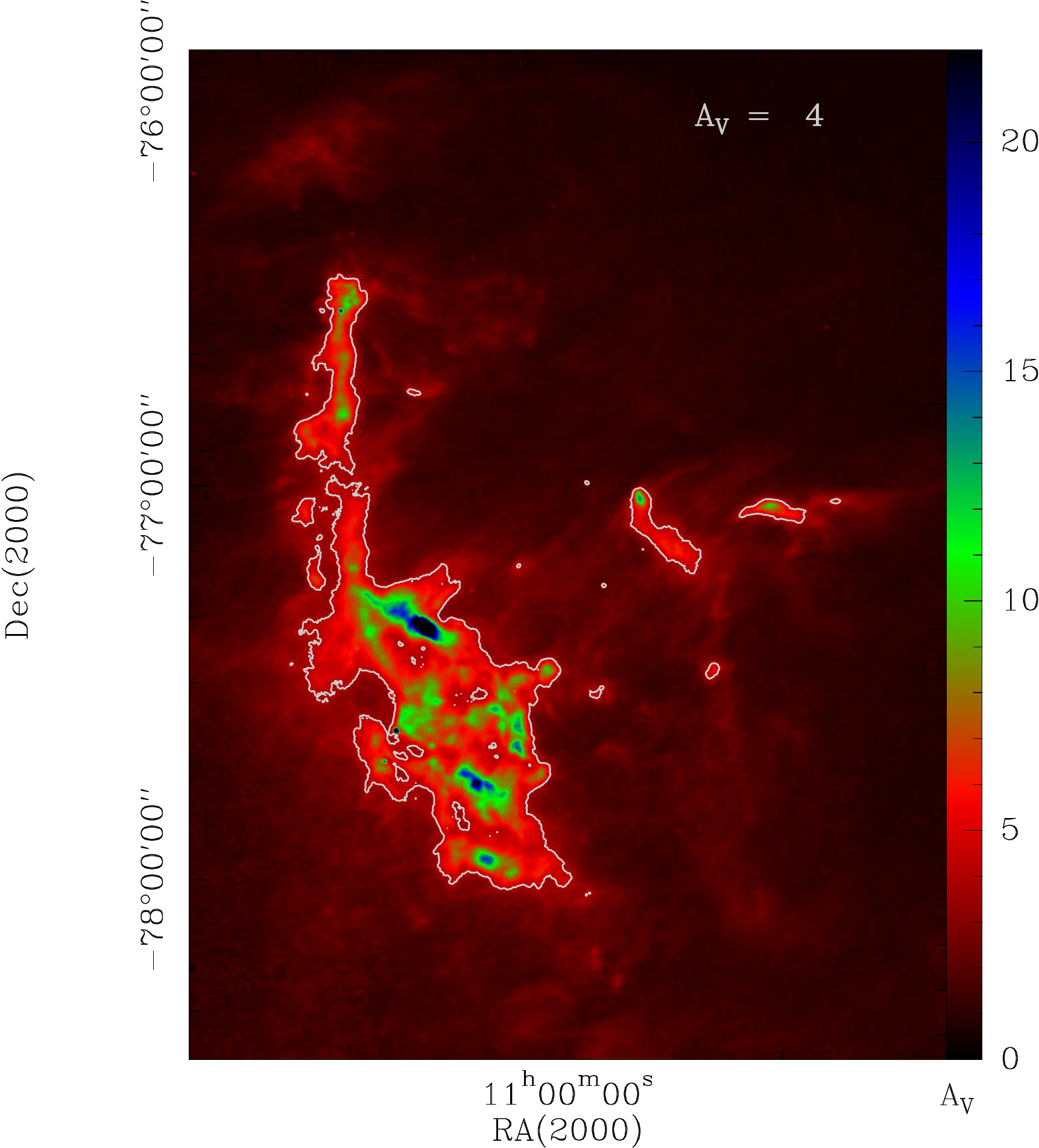}  
\caption{Column density map of Cha~I in which the colour scale is chosen to represent the characteristic $A_{V}$ levels (4, 10, and $\sim$20~mag) seen in the PDF. The $A_{V}$=4~mag level is indicated as a grey contour while the other two levels become obvious as a change in colour.}     
\label{cdmap2}  
\end{figure}  
  
Assuming our interpretation of a distinct, second (high)-density 
component for Cha~I is correct, the origin of this component remains 
to be understood. Various authors proposed that the star formation 
process is close to an end in Cha~I \citep[e.g.,][]{Belloche2011a}. 
Our \emph{Herschel} observations show that the region consists mainly 
of individual dense ($A_{V}>$10~mag) cores and clumps (size scales 0.2 
to 0.5~pc) embedded in a ridge of emission on a level of 
$A_{V}$=4--5~mag with a sharp border to the low-density regime that 
consists of \emph{striations} (see Figs.~\ref{spire} and \ref{cdmap2}, 
where the ridge is clearly outlined by the $A_{V}$ =4~mag contour). At 
this stage of evolution of the Cha~I cloud, mass transport to the 
ridge via \emph{striations} is unlikely to play a significant role if the 
star-formation process indeed has come to an end, in contrast to the 
Taurus B211 filament where \citet{Palmeirim2013} argued for mass 
supply to filaments. The faint \emph{striations} seen in Cha~I may be the 
left-over low-density gas, mixed with dust, aligned with the magnetic 
field lines \citep[][see Sect.~\ref{history}]{Whittet1994}. 
\citet{Kainulainen2011} proposed that such sharp spatial/density 
transition between a ridge and extended emission may be explained by 
pressure equilibrium. The authors interpreted the deviation of several 
observed PDFs from a lognormal -- turbulence dominated -- distribution 
at $A_{V}\approx4$ as a phase transition between a dense population of 
clumps and the diffuse interclump medium. This implies that for Cha~I, 
the whole ridge should be pressure confined with 
$<N_{ridge}><T_{ridge}>\approx<N_{ext}><T_{ext}>$ 
(assuming equal areas). However, the average external column density 
($<N>$) is at least a factor of 2 smaller than the one in the 
ridge, while the temperatures do not differ much 
($T_{ridge}$$\sim$12~K and $T_{ext}$$\sim$13~K, clearly seen in the 
\emph{striations}, and up to 14~K at a few pc distance). Pressure confinement 
would require much higher temperatures which is not observed. 
  
\subsection{The star-formation history in the Chamaeleon clouds} \label{history}   
In the following, and taking into account the ensemble of parameters extracted from the new {\sl Herschel} observations and previous studies of the Chamaelon clouds, we summarise the information on the star formation history across the three clouds. Cha~I has formed stars (clustered in the three Cederblad groups) and shows indications that it has arrived to its end of star-formation. This molecular cloud forms a ridge with a few filamentary structures and clumps. The ridge is embedded in low-density gas with \emph{striations} that run parallel with the magnetic field. Cha~II has a clumpy structure with ongoing star-formation, and faint \emph{striations} are only marginally observed. Cha~III shows no clear signs of star-formation, it's organised into a complex network of filaments with low column densities, and shows fewer structures that resemble \emph{striations}, particularly absent on the east part of the cloud.   
  
Assuming that the three clouds are part of a coherent complex (very likely in view of the $^{12}$CO data), the question arises as why they evolved in a different way. It is proposed by theory that molecular cloud formation is governed by large-scale turbulence such as expanding shock fronts due to supernovae explosions. The Chamaeleon region is indeed exposed to the Sco-Cen OB associations, and their stellar winds and supernovae might have initiated the formation of the clouds, which per se is not a reason why the clouds would end up with different properties due to this energy injection. Different initial distributions of atomic hydrogen and statistical dynamics (as it is seen in colliding \ion{H}{i} flow simulations) could have led to the observed cloud pattern, with filaments in Cha~III still evolving either to merge and built up a larger structure or dissolve, and a ridge-structure in Cha~I where the same process already came to end.

The magnetic field in the region could also play a significant role in shaping the cloud structure and star-formation activity. If we assume that the orientation of the large scale magnetic field measured for Cha~I does not change significantly towards Cha~III, the fewer \emph{striations} seen in this cloud do not appear to show a preferred alignment. Preliminary results from \emph{Planck}'s Galatic observations and MHD simulations, point towards an anti-correlation between the measured degree of polarisation and the dispersion of polarisation angles\footnote{http://www.rssd.esa.int/index.php?project=PLANCK\&page=47$\_$eslab}. In Chamaeleon, polarisation measurements from the \emph{Planck} satellite could be particularly relevant to understand their differences in star-formation history. Given that the large-scale turbulent structure of the three clouds does not vary by much across the regions, and the observed differences in the current density structure arise mainly from gravity, one could speculate that the initial difference in mass accretion along these \emph{striations} -- guided by the local magnetic field -- shaped the star-formation history of the three regions. This could explain why Cha~III is a quiescent cloud if it has not experienced this mass accretion process.
  
\section{Conclusions} \label{conclusion}  
 New \emph{Herschel} photometric observations of the three Chamaeleon clouds (I, II, and III) taken with the PACS and SPIRE instruments in an homogeneous way were presented in this paper. They were analysed with a set of tools to characterise quantitatively the large scale structure and extended dust emission, and study their possible relation to the accentuated differences seen today amongst the three clouds, with Cha~I likely at the end of star-formation, Cha~II actively forming stars, and Cha~III in a quiescent state.   
  
The column density and temperature maps derived from the \emph{Herschel} data reveal important morphological differences for the three clouds, with a ridge-like structure for Cha~I surrounded by faint filaments (\emph{striations}) aligned with the large-scale magnetic field, a clump-dominated regime for Cha~II, and a complex low-density filamentary network for Cha~III.   
  
Filamentary-like structures share a common width ($\sim$0.12$\pm$0.04~pc) consistent with values inferred from observations of other star-forming regions \citep[e.g.,][]{Arzoumanian2011,Andre2014}. However, only in Cha~I and II filaments are found to be predominately gravitationally unstable.   
  
All regions show a PDF described by a lognormal distribution for low column densities with a width of $\sim$0.45 to 0.6 mag and a peak at A$_{V}$$\sim$2~mag. For higher column densities, the PDFs show significant differences, with Cha~II being the only region where a classical single power-law tail with a slope indicative of free-fall collapse is seen. We compared the PDFs to the results from hydrodynamic simulations of \citet{Federrath2012}, and conclude that they are broadly described by models where $\mathcal{M}$=3, 5, or 10 with solenoidal or mixed forcing (with or without magnetic field) and low SFE.  
 
Overall, the turbulence properties of the three regions do not show large differences, lending strength to a scenario where the clouds are impacted by common large-scale processes. We emphasise, however, that an alignment of faint filaments peripheral to dense structures with the magnetic field is clearly seen in Cha~I. Similar preferential distributions have been found in other star-forming regions \citep[e.g.,][]{Palmeirim2013}. Future results from the \emph{Planck} mission on polarisation will quantify the relation between the magnetic filed and the structures in Cha~II and Cha~III. If proven that \emph{striations} are important channels of accretion of ambient material into filaments, the magnetic field could play an important role in shaping the differences in the star-formation history of the Chamaeleon complex regions.  
   
\begin{acknowledgements}  
  We thank the referee for a helpful review. We acknowledge support from the Faculty of the European Space 
  Astronomy Centre (ESAC/ESA). We thank Arnaud Belloche, and Pedro Palmeirim for their feedback on an earlier draft of this work. V. K. and Ph. A. acknowledge 
support from the European Research Council under the European Union's Seventh 
Framework Programme (FP7/2007-2013 -- ERC Grant Agreement no. 291294). N.S., S.B., and P.A. 
  acknowledge support by the ANR (\emph{Agence Nationale pour la 
    Recherche}) project ``STARFICH'', number ANR-11-BS56-010. N. 
  Schneider acknowledges funding by the DFG-priority program 1573 
  (ISM-SPP). AR is funded by the ESAC Science Operations Division 
  research funds with code SC 1300016149.  NLJC acknowledges support 
  from the Belgian Federal Science Policy Office via the PRODEX 
  Programme of ESA. SPIRE has been developed by a consortium of 
  institutes led by Cardiff Univ. (UK) and including: Univ. Lethbridge 
  (Canada); NAOC (China); CEA, LAM (France); IFSI, Univ. Padua 
  (Italy); IAC (Spain); Stockholm Observatory (Sweden); Imperial 
  College London, RAL, UCL-MSSL, UKATC, Univ. Sussex (UK); and 
  Caltech, JPL, NHSC, Univ. Colorado (USA).  This development has been 
  supported by national funding agencies: CSA (Canada); NAOC (China); 
  CEA, CNES, CNRS (France); ASI (Italy); MCINN (Spain); SNSB (Sweden); 
  STFC, UKSA (UK); and NASA (USA). PACS has been developed by a 
  consortium of institutes led by MPE (Germany) and including UVIE 
  (Austria); KU Leuven, CSL, IMEC (Belgium); CEA, LAM (France); MPIA 
  (Germany); INAF- IFSI/OAA/OAP/OAT, LENS, SISSA (Italy); IAC (Spain). 
  This development has been supported by the funding agencies BMVIT 
  (Austria), ESA-PRODEX (Belgium), CEA/CNES (France), DLR (Germany), 
  ASI/INAF (Italy), and CICYT/MCYT (Spain).

\end{acknowledgements}


\begin{thebibliography}{64}
\expandafter\ifx\csname natexlab\endcsname\relax\def\natexlab#1{#1}\fi

\bibitem[{{Andr{\'e}} {et~al.}(2014){Andr{\'e}}, {Di Francesco},
  {Ward-Thompson}, {Inutsuka}, {Pudritz}, \& {Pineda}}]{Andre2014}
{Andr{\'e}}, P., {Di Francesco}, J., {Ward-Thompson}, D., {et~al.} 2014,
  Protostars and Planets VI, in press (arXiv:astro-ph/1312.6232)

\bibitem[{{Andr{\'e}} {et~al.}(2010){Andr{\'e}}, {Men'shchikov}, {Bontemps},
  {K{\"o}nyves}, {Motte}, {Schneider}, {Didelon}, {Minier}, {Saraceno},
  {Ward-Thompson}, {di Francesco}, {White}, {Molinari}, {Testi}, {Abergel},
  {Griffin}, {Henning}, {Royer}, {Mer{\'{\i}}n}, {Vavrek}, {Attard},
  {Arzoumanian}, {Wilson}, {Ade}, {Aussel}, {Baluteau}, {Benedettini},
  {Bernard}, {Blommaert}, {Cambr{\'e}sy}, {Cox}, {di Giorgio}, {Hargrave},
  {Hennemann}, {Huang}, {Kirk}, {Krause}, {Launhardt}, {Leeks}, {Le Pennec},
  {Li}, {Martin}, {Maury}, {Olofsson}, {Omont}, {Peretto}, {Pezzuto}, {Prusti},
  {Roussel}, {Russeil}, {Sauvage}, {Sibthorpe}, {Sicilia-Aguilar}, {Spinoglio},
  {Waelkens}, {Woodcraft}, \& {Zavagno}}]{Andre2010}
{Andr{\'e}}, P., {Men'shchikov}, A., {Bontemps}, S., {et~al.} 2010, \aap, 518,
  L102

\bibitem[{{Arzoumanian} {et~al.}(2011){Arzoumanian}, {Andr{\'e}}, {Didelon},
  {K{\"o}nyves}, {Schneider}, {Men'shchikov}, {Sousbie}, {Zavagno}, {Bontemps},
  {di Francesco}, {Griffin}, {Hennemann}, {Hill}, {Kirk}, {Martin}, {Minier},
  {Molinari}, {Motte}, {Peretto}, {Pezzuto}, {Spinoglio}, {Ward-Thompson},
  {White}, \& {Wilson}}]{Arzoumanian2011}
{Arzoumanian}, D., {Andr{\'e}}, P., {Didelon}, P., {et~al.} 2011, \aap, 529, L6

\bibitem[{{Ballesteros-Paredes} {et~al.}(2011){Ballesteros-Paredes},
  {V{\'a}zquez-Semadeni}, {Gazol}, {Hartmann}, {Heitsch}, \&
  {Col{\'{\i}}n}}]{Ballesteros-Paredes2011}
{Ballesteros-Paredes}, J., {V{\'a}zquez-Semadeni}, E., {Gazol}, A., {et~al.}
  2011, \mnras, 416, 1436

\bibitem[{{Belloche} {et~al.}(2011{\natexlab{a}}){Belloche}, {Parise},
  {Schuller}, {Andr{\'e}}, {Bontemps}, \& {Menten}}]{Belloche2011b}
{Belloche}, A., {Parise}, B., {Schuller}, F., {et~al.} 2011{\natexlab{a}},
  \aap, 535, A2

\bibitem[{{Belloche} {et~al.}(2011{\natexlab{b}}){Belloche}, {Schuller},
  {Parise}, {Andr{\'e}}, {Hatchell}, {J{\o}rgensen}, {Bontemps}, {Wei{\ss}},
  {Menten}, \& {Muders}}]{Belloche2011a}
{Belloche}, A., {Schuller}, F., {Parise}, B., {et~al.} 2011{\natexlab{b}},
  \aap, 527, A145

\bibitem[{{Bensch} {et~al.}(2001){Bensch}, {Stutzki}, \&
  {Ossenkopf}}]{Bensch2001}
{Bensch}, F., {Stutzki}, J., \& {Ossenkopf}, V. 2001, \aap, 366, 636

\bibitem[{{Bernard} {et~al.}(2010){Bernard}, {Paradis}, {Marshall}, {Montier},
  {Lagache}, {Paladini}, {Veneziani}, {Brunt}, {Mottram}, {Martin},
  {Ristorcelli}, {Noriega-Crespo}, {Compi{\`e}gne}, {Flagey}, {Anderson},
  {Popescu}, {Tuffs}, {Reach}, {White}, {Benedetti}, {Calzoletti}, {Digiorgio},
  {Faustini}, {Juvela}, {Joblin}, {Joncas}, {Mivilles-Deschenes}, {Olmi},
  {Traficante}, {Piacentini}, {Zavagno}, \& {Molinari}}]{Bernard2010}
{Bernard}, J.-P., {Paradis}, D., {Marshall}, D.~J., {et~al.} 2010, \aap, 518,
  L88

\bibitem[{{Bohlin} {et~al.}(1978){Bohlin}, {Savage}, \& {Drake}}]{Bohlin1978}
{Bohlin}, R.~C., {Savage}, B.~D., \& {Drake}, J.~F. 1978, \apj, 224, 132

\bibitem[{{Brunt} {et~al.}(2010){Brunt}, {Federrath}, \& {Price}}]{Brunt2010}
{Brunt}, C.~M., {Federrath}, C., \& {Price}, D.~J. 2010, \mnras, 405, L56

\bibitem[{{Federrath} \& {Klessen}(2012)}]{Federrath2012}
{Federrath}, C. \& {Klessen}, R.~S. 2012, \apj, 761, 156

\bibitem[{{Federrath} \& {Klessen}(2013)}]{Federrath2013}
{Federrath}, C. \& {Klessen}, R.~S. 2013, \apj, 763, 51

\bibitem[{{Federrath} {et~al.}(2010){Federrath}, {Roman-Duval}, {Klessen},
  {Schmidt}, \& {Mac Low}}]{Federrath2010}
{Federrath}, C., {Roman-Duval}, J., {Klessen}, R.~S., {Schmidt}, W., \& {Mac
  Low}, M.-M. 2010, \aap, 512, A81

\bibitem[{{Federrath} {et~al.}(2011){Federrath}, {Sur}, {Schleicher},
  {Banerjee}, \& {Klessen}}]{Federrath2011}
{Federrath}, C., {Sur}, S., {Schleicher}, D.~R.~G., {Banerjee}, R., \&
  {Klessen}, R.~S. 2011, \apj, 731, 62

\bibitem[{{Froebrich} \& {Rowles}(2010)}]{Froebrich2010}
{Froebrich}, D. \& {Rowles}, J. 2010, \mnras, 406, 1350

\bibitem[{{Girichidis} {et~al.}(2014){Girichidis}, {Konstandin}, {Whitworth},
  \& {Klessen}}]{Girichidis2014}
{Girichidis}, P., {Konstandin}, L., {Whitworth}, A.~P., \& {Klessen}, R.~S.
  2014, \apj, 781, 91

\bibitem[{{Griffin} {et~al.}(2010){Griffin}, {Abergel}, {Abreu}, {Ade},
  {Andr{\'e}}, {Augueres}, {Babbedge}, {Bae}, {Baillie}, {Baluteau}, {Barlow},
  {Bendo}, {Benielli}, {Bock}, {Bonhomme}, {Brisbin}, {Brockley-Blatt},
  {Caldwell}, {Cara}, {Castro-Rodriguez}, {Cerulli}, {Chanial}, {Chen},
  {Clark}, {Clements}, {Clerc}, {Coker}, {Communal}, {Conversi}, {Cox},
  {Crumb}, {Cunningham}, {Daly}, {Davis}, {de Antoni}, {Delderfield}, {Devin},
  {di Giorgio}, {Didschuns}, {Dohlen}, {Donati}, {Dowell}, {Dowell}, {Duband},
  {Dumaye}, {Emery}, {Ferlet}, {Ferrand}, {Fontignie}, {Fox}, {Franceschini},
  {Frerking}, {Fulton}, {Garcia}, {Gastaud}, {Gear}, {Glenn}, {Goizel},
  {Griffin}, {Grundy}, {Guest}, {Guillemet}, {Hargrave}, {Harwit}, {Hastings},
  {Hatziminaoglou}, {Herman}, {Hinde}, {Hristov}, {Huang}, {Imhof}, {Isaak},
  {Israelsson}, {Ivison}, {Jennings}, {Kiernan}, {King}, {Lange}, {Latter},
  {Laurent}, {Laurent}, {Leeks}, {Lellouch}, {Levenson}, {Li}, {Li},
  {Lilienthal}, {Lim}, {Liu}, {Lu}, {Madden}, {Mainetti}, {Marliani}, {McKay},
  {Mercier}, {Molinari}, {Morris}, {Moseley}, {Mulder}, {Mur}, {Naylor},
  {Nguyen}, {O'Halloran}, {Oliver}, {Olofsson}, {Olofsson}, {Orfei}, {Page},
  {Pain}, {Panuzzo}, {Papageorgiou}, {Parks}, {Parr-Burman}, {Pearce},
  {Pearson}, {P{\'e}rez-Fournon}, {Pinsard}, {Pisano}, {Podosek}, {Pohlen},
  {Polehampton}, {Pouliquen}, {Rigopoulou}, {Rizzo}, {Roseboom}, {Roussel},
  {Rowan-Robinson}, {Rownd}, {Saraceno}, {Sauvage}, {Savage}, {Savini},
  {Sawyer}, {Scharmberg}, {Schmitt}, {Schneider}, {Schulz}, {Schwartz},
  {Shafer}, {Shupe}, {Sibthorpe}, {Sidher}, {Smith}, {Smith}, {Smith},
  {Spencer}, {Stobie}, {Sudiwala}, {Sukhatme}, {Surace}, {Stevens}, {Swinyard},
  {Trichas}, {Tourette}, {Triou}, {Tseng}, {Tucker}, {Turner}, {Vaccari},
  {Valtchanov}, {Vigroux}, {Virique}, {Voellmer}, {Walker}, {Ward}, {Waskett},
  {Weilert}, {Wesson}, {White}, {Whitehouse}, {Wilson}, {Winter}, {Woodcraft},
  {Wright}, {Xu}, {Zavagno}, {Zemcov}, {Zhang}, \& {Zonca}}]{Griffin2010}
{Griffin}, M.~J., {Abergel}, A., {Abreu}, A., {et~al.} 2010, \aap, 518, L3

\bibitem[{{Hacar} {et~al.}(2013){Hacar}, {Tafalla}, {Kauffmann}, \&
  {Kov{\'a}cs}}]{Hacar2013}
{Hacar}, A., {Tafalla}, M., {Kauffmann}, J., \& {Kov{\'a}cs}, A. 2013, \aap,
  554, A55

\bibitem[{{Haikala} {et~al.}(2005){Haikala}, {Harju}, {Mattila}, \&
  {Toriseva}}]{Haikala2005}
{Haikala}, L.~K., {Harju}, J., {Mattila}, K., \& {Toriseva}, M. 2005, \aap,
  431, 149

\bibitem[{{Hennebelle}(2013)}]{Hennebelle2013}
{Hennebelle}, P. 2013, \aap, 556, A153

\bibitem[{{Hennemann} {et~al.}(2012){Hennemann}, {Motte}, {Schneider},
  {Didelon}, {Hill}, {Arzoumanian}, {Bontemps}, {Csengeri}, {Andr{\'e}},
  {Konyves}, {Louvet}, {Marston}, {Men'shchikov}, {Minier}, {Nguyen Luong},
  {Palmeirim}, {Peretto}, {Sauvage}, {Zavagno}, {Anderson}, {Bernard}, {Di
  Francesco}, {Elia}, {Li}, {Martin}, {Molinari}, {Pezzuto}, {Russeil}, {Rygl},
  {Schisano}, {Spinoglio}, {Sousbie}, {Ward-Thompson}, \&
  {White}}]{Hennemann2012}
{Hennemann}, M., {Motte}, F., {Schneider}, N., {et~al.} 2012, \aap, 543, L3

\bibitem[{{Hildebrand}(1983)}]{Hildebrand1983}
{Hildebrand}, R.~H. 1983, \qjras, 24, 267

\bibitem[{{Kainulainen} {et~al.}(2011){Kainulainen}, {Beuther}, {Banerjee},
  {Federrath}, \& {Henning}}]{Kainulainen2011}
{Kainulainen}, J., {Beuther}, H., {Banerjee}, R., {Federrath}, C., \&
  {Henning}, T. 2011, \aap, 530, A64

\bibitem[{{Kainulainen} {et~al.}(2009){Kainulainen}, {Beuther}, {Henning}, \&
  {Plume}}]{Kainulainen2009}
{Kainulainen}, J., {Beuther}, H., {Henning}, T., \& {Plume}, R. 2009, \aap,
  508, L35

\bibitem[{{Klessen}(2000)}]{Klessen2000}
{Klessen}, R.~S. 2000, \apj, 535, 869

\bibitem[{{K{\"o}nyves} {et~al.}(2010){K{\"o}nyves}, {Andr{\'e}},
  {Men'shchikov}, {Schneider}, {Arzoumanian}, {Bontemps}, {Attard}, {Motte},
  {Didelon}, {Maury}, {Abergel}, {Ali}, {Baluteau}, {Bernard}, {Cambr{\'e}sy},
  {Cox}, {di Francesco}, {di Giorgio}, {Griffin}, {Hargrave}, {Huang}, {Kirk},
  {Li}, {Martin}, {Minier}, {Molinari}, {Olofsson}, {Pezzuto}, {Russeil},
  {Roussel}, {Saraceno}, {Sauvage}, {Sibthorpe}, {Spinoglio}, {Testi},
  {Ward-Thompson}, {White}, {Wilson}, {Woodcraft}, \& {Zavagno}}]{Konyves2010}
{K{\"o}nyves}, V., {Andr{\'e}}, P., {Men'shchikov}, A., {et~al.} 2010, \aap,
  518, L106

\bibitem[{{K{\'o}sp{\'a}l} {et~al.}(2012){K{\'o}sp{\'a}l}, {Prusti}, {Cox},
  {Pilbratt}, {Andr{\'e}}, {Alves de Oliveira}, {Winston}, {Mer{\'{\i}}n},
  {Ribas}, {Royer}, {Vavrek}, \& {Waelkens}}]{Kospal2012}
{K{\'o}sp{\'a}l}, {\'A}., {Prusti}, T., {Cox}, N.~L.~J., {et~al.} 2012, \aap,
  541, A71

\bibitem[{{Kritsuk} {et~al.}(2013){Kritsuk}, {Lee}, \& {Norman}}]{Kritsuk2013}
{Kritsuk}, A.~G., {Lee}, C.~T., \& {Norman}, M.~L. 2013, \mnras

\bibitem[{{Lombardi} {et~al.}(2008){Lombardi}, {Lada}, \&
  {Alves}}]{Lombardi2008}
{Lombardi}, M., {Lada}, C.~J., \& {Alves}, J. 2008, \aap, 489, 143

\bibitem[{{Luhman}(2008)}]{Luhman2008}
{Luhman}, K.~L. 2008, {Chamaeleon}, ed. B.~{Reipurth}, 169

\bibitem[{{Mac Low} \& {Ossenkopf}(2000)}]{MacLow2000}
{Mac Low}, M.-M. \& {Ossenkopf}, V. 2000, \aap, 353, 339

\bibitem[{{McGregor} {et~al.}(1994){McGregor}, {Harrison}, {Hough}, \&
  {Bailey}}]{McGregor1994}
{McGregor}, P.~J., {Harrison}, T.~E., {Hough}, J.~H., \& {Bailey}, J.~A. 1994,
  \mnras, 267, 755

\bibitem[{{Men'shchikov} {et~al.}(2010){Men'shchikov}, {Andr{\'e}}, {Didelon},
  {K{\"o}nyves}, {Schneider}, {Motte}, {Bontemps}, {Arzoumanian}, {Attard},
  {Abergel}, {Baluteau}, {Bernard}, {Cambr{\'e}sy}, {Cox}, {di Francesco}, {di
  Giorgio}, {Griffin}, {Hargrave}, {Huang}, {Kirk}, {Li}, {Martin}, {Minier},
  {Miville-Desch{\^e}nes}, {Molinari}, {Olofsson}, {Pezzuto}, {Roussel},
  {Russeil}, {Saraceno}, {Sauvage}, {Sibthorpe}, {Spinoglio}, {Testi},
  {Ward-Thompson}, {White}, {Wilson}, {Woodcraft}, \&
  {Zavagno}}]{Menshchikov2010}
{Men'shchikov}, A., {Andr{\'e}}, P., {Didelon}, P., {et~al.} 2010, \aap, 518,
  L103

\bibitem[{{Miville-Desch{\^e}nes} {et~al.}(2010){Miville-Desch{\^e}nes},
  {Martin}, {Abergel}, {Bernard}, {Boulanger}, {Lagache}, {Anderson},
  {Andr{\'e}}, {Arab}, {Baluteau}, {Blagrave}, {Bontemps}, {Cohen},
  {Compiegne}, {Cox}, {Dartois}, {Davis}, {Emery}, {Fulton}, {Gry}, {Habart},
  {Huang}, {Joblin}, {Jones}, {Kirk}, {Lim}, {Madden}, {Makiwa}, {Menshchikov},
  {Molinari}, {Moseley}, {Motte}, {Naylor}, {Okumura}, {Pinheiro Gon{\c
  c}alves}, {Polehampton}, {Rod{\'o}n}, {Russeil}, {Saraceno}, {Schneider},
  {Sidher}, {Spencer}, {Swinyard}, {Ward-Thompson}, {White}, \&
  {Zavagno}}]{Miville2010}
{Miville-Desch{\^e}nes}, M.-A., {Martin}, P.~G., {Abergel}, A., {et~al.} 2010,
  \aap, 518, L104

\bibitem[{{Mizuno} {et~al.}(2001){Mizuno}, {Yamaguchi}, {Tachihara}, {Toyoda},
  {Aoyama}, {Yamamoto}, {Onishi}, \& {Fukui}}]{Mizuno2001}
{Mizuno}, A., {Yamaguchi}, R., {Tachihara}, K., {et~al.} 2001, \pasj, 53, 1071

\bibitem[{{Molinari} {et~al.}(2010){Molinari}, {Swinyard}, {Bally}, {Barlow},
  {Bernard}, {Martin}, {Moore}, {Noriega-Crespo}, {Plume}, {Testi}, {Zavagno},
  {Abergel}, {Ali}, {Anderson}, {Andr{\'e}}, {Baluteau}, {Battersby},
  {Beltr{\'a}n}, {Benedettini}, {Billot}, {Blommaert}, {Bontemps}, {Boulanger},
  {Brand}, {Brunt}, {Burton}, {Calzoletti}, {Carey}, {Caselli}, {Cesaroni},
  {Cernicharo}, {Chakrabarti}, {Chrysostomou}, {Cohen}, {Compiegne}, {de
  Bernardis}, {de Gasperis}, {di Giorgio}, {Elia}, {Faustini}, {Flagey},
  {Fukui}, {Fuller}, {Ganga}, {Garcia-Lario}, {Glenn}, {Goldsmith}, {Griffin},
  {Hoare}, {Huang}, {Ikhenaode}, {Joblin}, {Joncas}, {Juvela}, {Kirk},
  {Lagache}, {Li}, {Lim}, {Lord}, {Marengo}, {Marshall}, {Masi}, {Massi},
  {Matsuura}, {Minier}, {Miville-Desch{\^e}nes}, {Montier}, {Morgan}, {Motte},
  {Mottram}, {M{\"u}ller}, {Natoli}, {Neves}, {Olmi}, {Paladini}, {Paradis},
  {Parsons}, {Peretto}, {Pestalozzi}, {Pezzuto}, {Piacentini}, {Piazzo},
  {Polychroni}, {Pomar{\`e}s}, {Popescu}, {Reach}, {Ristorcelli}, {Robitaille},
  {Robitaille}, {Rod{\'o}n}, {Roy}, {Royer}, {Russeil}, {Saraceno}, {Sauvage},
  {Schilke}, {Schisano}, {Schneider}, {Schuller}, {Schulz}, {Sibthorpe},
  {Smith}, {Smith}, {Spinoglio}, {Stamatellos}, {Strafella}, {Stringfellow},
  {Sturm}, {Taylor}, {Thompson}, {Traficante}, {Tuffs}, {Umana}, {Valenziano},
  {Vavrek}, {Veneziani}, {Viti}, {Waelkens}, {Ward-Thompson}, {White},
  {Wilcock}, {Wyrowski}, {Yorke}, \& {Zhang}}]{Molinari2010}
{Molinari}, S., {Swinyard}, B., {Bally}, J., {et~al.} 2010, \aap, 518, L100

\bibitem[{{Motte} {et~al.}(2010){Motte}, {Zavagno}, {Bontemps}, {Schneider},
  {Hennemann}, {di Francesco}, {Andr{\'e}}, {Saraceno}, {Griffin}, {Marston},
  {Ward-Thompson}, {White}, {Minier}, {Men'shchikov}, {Hill}, {Abergel},
  {Anderson}, {Aussel}, {Balog}, {Baluteau}, {Bernard}, {Cox}, {Csengeri},
  {Deharveng}, {Didelon}, {di Giorgio}, {Hargrave}, {Huang}, {Kirk}, {Leeks},
  {Li}, {Martin}, {Molinari}, {Nguyen-Luong}, {Olofsson}, {Persi}, {Peretto},
  {Pezzuto}, {Roussel}, {Russeil}, {Sadavoy}, {Sauvage}, {Sibthorpe},
  {Spinoglio}, {Testi}, {Teyssier}, {Vavrek}, {Wilson}, \&
  {Woodcraft}}]{Motte2010}
{Motte}, F., {Zavagno}, A., {Bontemps}, S., {et~al.} 2010, \aap, 518, L77

\bibitem[{{Ossenkopf} {et~al.}(2008{\natexlab{a}}){Ossenkopf}, {Krips}, \&
  {Stutzki}}]{Ossenkopf2008a}
{Ossenkopf}, V., {Krips}, M., \& {Stutzki}, J. 2008{\natexlab{a}}, \aap, 485,
  917

\bibitem[{{Ossenkopf} {et~al.}(2008{\natexlab{b}}){Ossenkopf}, {Krips}, \&
  {Stutzki}}]{Ossenkopf2008b}
{Ossenkopf}, V., {Krips}, M., \& {Stutzki}, J. 2008{\natexlab{b}}, \aap, 485,
  719

\bibitem[{{Ott}(2010)}]{Ott2010}
{Ott}, S. 2010, in Astronomical Society of the Pacific Conference Series, Vol.
  434, Astronomical Data Analysis Software and Systems XIX, ed. Y.~{Mizumoto},
  K.-I. {Morita}, \& M.~{Ohishi}, 139

\bibitem[{{Palmeirim} {et~al.}(2013){Palmeirim}, {Andr{\'e}}, {Kirk},
  {Ward-Thompson}, {Arzoumanian}, {K{\"o}nyves}, {Didelon}, {Schneider},
  {Benedettini}, {Bontemps}, {Di Francesco}, {Elia}, {Griffin}, {Hennemann},
  {Hill}, {Martin}, {Men'shchikov}, {Molinari}, {Motte}, {Nguyen Luong},
  {Nutter}, {Peretto}, {Pezzuto}, {Roy}, {Rygl}, {Spinoglio}, \&
  {White}}]{Palmeirim2013}
{Palmeirim}, P., {Andr{\'e}}, P., {Kirk}, J., {et~al.} 2013, \aap, 550, A38

\bibitem[{{Passot} \& {V{\'a}zquez-Semadeni}(1998)}]{Passot1998}
{Passot}, T. \& {V{\'a}zquez-Semadeni}, E. 1998, \pre, 58, 4501

\bibitem[{{Peretto} {et~al.}(2013){Peretto}, {Fuller}, {Duarte-Cabral},
  {Avison}, {Hennebelle}, {Pineda}, {Andr{\'e}}, {Bontemps}, {Motte},
  {Schneider}, \& {Molinari}}]{Peretto2013}
{Peretto}, N., {Fuller}, G.~A., {Duarte-Cabral}, A., {et~al.} 2013, \aap, 555,
  A112

\bibitem[{{Pilbratt} {et~al.}(2010){Pilbratt}, {Riedinger}, {Passvogel},
  {Crone}, {Doyle}, {Gageur}, {Heras}, {Jewell}, {Metcalfe}, {Ott}, \&
  {Schmidt}}]{Pilbratt2010}
{Pilbratt}, G.~L., {Riedinger}, J.~R., {Passvogel}, T., {et~al.} 2010, \aap,
  518, L1

\bibitem[{{Pineda} {et~al.}(2008){Pineda}, {Caselli}, \&
  {Goodman}}]{Pineda2008}
{Pineda}, J.~E., {Caselli}, P., \& {Goodman}, A.~A. 2008, \apj, 679, 481

\bibitem[{{Poglitsch} {et~al.}(2010){Poglitsch}, {Waelkens}, {Geis},
  {Feuchtgruber}, {Vandenbussche}, {Rodriguez}, {Krause}, {Renotte}, {van
  Hoof}, {Saraceno}, {Cepa}, {Kerschbaum}, {Agn{\`e}se}, {Ali}, {Altieri},
  {Andreani}, {Augueres}, {Balog}, {Barl}, {Bauer}, {Belbachir}, {Benedettini},
  {Billot}, {Boulade}, {Bischof}, {Blommaert}, {Callut}, {Cara}, {Cerulli},
  {Cesarsky}, {Contursi}, {Creten}, {De Meester}, {Doublier}, {Doumayrou},
  {Duband}, {Exter}, {Genzel}, {Gillis}, {Gr{\"o}zinger}, {Henning},
  {Herreros}, {Huygen}, {Inguscio}, {Jakob}, {Jamar}, {Jean}, {de Jong},
  {Katterloher}, {Kiss}, {Klaas}, {Lemke}, {Lutz}, {Madden}, {Marquet},
  {Martignac}, {Mazy}, {Merken}, {Montfort}, {Morbidelli}, {M{\"u}ller},
  {Nielbock}, {Okumura}, {Orfei}, {Ottensamer}, {Pezzuto}, {Popesso},
  {Putzeys}, {Regibo}, {Reveret}, {Royer}, {Sauvage}, {Schreiber}, {Stegmaier},
  {Schmitt}, {Schubert}, {Sturm}, {Thiel}, {Tofani}, {Vavrek}, {Wetzstein},
  {Wieprecht}, \& {Wiezorrek}}]{Poglitsch2010}
{Poglitsch}, A., {Waelkens}, C., {Geis}, N., {et~al.} 2010, \aap, 518, L2

\bibitem[{{Polychroni} {et~al.}(2013){Polychroni}, {Schisano}, {Elia}, {Roy},
  {Molinari}, {Martin}, {Andr{\'e}}, {Turrini}, {Rygl}, {Di Francesco},
  {Benedettini}, {Busquet}, {di Giorgio}, {Pestalozzi}, {Pezzuto},
  {Arzoumanian}, {Bontemps}, {Hennemann}, {Hill}, {K{\"o}nyves},
  {Men'shchikov}, {Motte}, {Nguyen-Luong}, {Peretto}, {Schneider}, \&
  {White}}]{Polychroni2013}
{Polychroni}, D., {Schisano}, E., {Elia}, D., {et~al.} 2013, \apjl, 777, L33

\bibitem[{{Rivera-Ingraham} {et~al.}(2013){Rivera-Ingraham}, {Martin},
  {Polychroni}, {Motte}, {Schneider}, {Bontemps}, {Hennemann}, {Men'shchikov},
  {Nguyen Luong}, {Andr{\'e}}, {Arzoumanian}, {Bernard}, {Di Francesco},
  {Elia}, {Fallscheer}, {Hill}, {Li}, {Minier}, {Pezzuto}, {Roy}, {Rygl},
  {Sadavoy}, {Spinoglio}, {White}, \& {Wilson}}]{Rivera-Ingraham2013}
{Rivera-Ingraham}, A., {Martin}, P.~G., {Polychroni}, D., {et~al.} 2013, \apj,
  766, 85

\bibitem[{{Roussel}(2013)}]{Roussel2013}
{Roussel}, H. 2013, \pasp, 125, 1126

\bibitem[{{Roy} {et~al.}(2014){Roy}, {Andr{\'e}}, {Palmeirim}, {Attard},
  {K{\"o}nyves}, {Schneider}, {Peretto}, {Men'shchikov}, {Ward-Thompson},
  {Kirk}, {Griffin}, {Marsh}, {Abergel}, {Arzoumanian}, {Benedettini}, {Hill},
  {Motte}, {Nguyen Luong}, {Pezzuto}, {Rivera-Ingraham}, {Roussel}, {Rygl},
  {Spinoglio}, {Stamatellos}, \& {White}}]{Roy2014}
{Roy}, A., {Andr{\'e}}, P., {Palmeirim}, P., {et~al.} 2014, \aap, 562, A138

\bibitem[{{Schneider} {et~al.}(2013){Schneider}, {Andr{\'e}}, {K{\"o}nyves},
  {Bontemps}, {Motte}, {Federrath}, {Ward-Thompson}, {Arzoumanian},
  {Benedettini}, {Bressert}, {Didelon}, {Di Francesco}, {Griffin}, {Hennemann},
  {Hill}, {Palmeirim}, {Pezzuto}, {Peretto}, {Roy}, {Rygl}, {Spinoglio}, \&
  {White}}]{Schneider2013}
{Schneider}, N., {Andr{\'e}}, P., {K{\"o}nyves}, V., {et~al.} 2013, \apjl, 766

\bibitem[{{Schneider} {et~al.}(2011){Schneider}, {Bontemps}, {Simon},
  {Ossenkopf}, {Federrath}, {Klessen}, {Motte}, {Andr{\'e}}, {Stutzki}, \&
  {Brunt}}]{Schneider2011}
{Schneider}, N., {Bontemps}, S., {Simon}, R., {et~al.} 2011, \aap, 529, A1

\bibitem[{{Schneider} {et~al.}(2010){Schneider}, {Csengeri}, {Bontemps},
  {Motte}, {Simon}, {Hennebelle}, {Federrath}, \& {Klessen}}]{Schneider2010}
{Schneider}, N., {Csengeri}, T., {Bontemps}, S., {et~al.} 2010, \aap, 520, A49

\bibitem[{{Schneider} {et~al.}(2012){Schneider}, {Csengeri}, {Hennemann},
  {Motte}, {Didelon}, {Federrath}, {Bontemps}, {Di Francesco}, {Arzoumanian},
  {Minier}, {Andr{\'e}}, {Hill}, {Zavagno}, {Nguyen-Luong}, {Attard},
  {Bernard}, {Elia}, {Fallscheer}, {Griffin}, {Kirk}, {Klessen}, {K{\"o}nyves},
  {Martin}, {Men'shchikov}, {Palmeirim}, {Peretto}, {Pestalozzi}, {Russeil},
  {Sadavoy}, {Sousbie}, {Testi}, {Tremblin}, {Ward-Thompson}, \&
  {White}}]{Schneider2012}
{Schneider}, N., {Csengeri}, T., {Hennemann}, M., {et~al.} 2012, \aap, 540, L11

\bibitem[{{Shu} {et~al.}(1987){Shu}, {Adams}, \& {Lizano}}]{Shu1987}
{Shu}, F.~H., {Adams}, F.~C., \& {Lizano}, S. 1987, \araa, 25, 23

\bibitem[{{Sousbie}(2011)}]{Sousbie2011a}
{Sousbie}, T. 2011, \mnras, 414, 350

\bibitem[{{Sousbie} {et~al.}(2011){Sousbie}, {Pichon}, \&
  {Kawahara}}]{Sousbie2011b}
{Sousbie}, T., {Pichon}, C., \& {Kawahara}, H. 2011, \mnras, 414, 384

\bibitem[{{Spezzi} {et~al.}(2008){Spezzi}, {Alcal{\'a}}, {Covino}, {Frasca},
  {Gandolfi}, {Oliveira}, {Chapman}, {Evans}, {Huard}, {J{\o}rgensen},
  {Mer{\'{\i}}n}, \& {Stapelfeldt}}]{Spezzi2008}
{Spezzi}, L., {Alcal{\'a}}, J.~M., {Covino}, E., {et~al.} 2008, \apj, 680, 1295

\bibitem[{{Spezzi} {et~al.}(2013){Spezzi}, {Cox}, {Prusti}, {Mer{\'{\i}}n},
  {Ribas}, {Alves de Oliveira}, {Winston}, {K{\'o}sp{\'a}l}, {Royer}, {Vavrek},
  {Andr{\'e}}, {Pilbratt}, {Testi}, {Bressert}, {Ricci}, {Men'shchikov}, \&
  {K{\"o}nyves}}]{Spezzi2013}
{Spezzi}, L., {Cox}, N.~L.~J., {Prusti}, T., {et~al.} 2013, \aap, 555, A71

\bibitem[{{Starck} {et~al.}(2003){Starck}, {Donoho}, \&
  {Cand{\`e}s}}]{Starck2003}
{Starck}, J.~L., {Donoho}, D.~L., \& {Cand{\`e}s}, E.~J. 2003, \aap, 398, 785

\bibitem[{{Stutzki} {et~al.}(1998){Stutzki}, {Bensch}, {Heithausen},
  {Ossenkopf}, \& {Zielinsky}}]{Stutzki1998}
{Stutzki}, J., {Bensch}, F., {Heithausen}, A., {Ossenkopf}, V., \& {Zielinsky},
  M. 1998, \aap, 336, 697

\bibitem[{{Whittet} {et~al.}(1994){Whittet}, {Gerakines}, {Carkner}, {Hough},
  {Martin}, {Prusti}, \& {Kilkenny}}]{Whittet1994}
{Whittet}, D.~C.~B., {Gerakines}, P.~A., {Carkner}, A.~L., {et~al.} 1994,
  \mnras, 268, 1

\bibitem[{{Whittet} {et~al.}(1997){Whittet}, {Prusti}, {Franco}, {Gerakines},
  {Kilkenny}, {Larson}, \& {Wesselius}}]{Whittet1997}
{Whittet}, D.~C.~B., {Prusti}, T., {Franco}, G.~A.~P., {et~al.} 1997, \aap,
  327, 1194

\bibitem[{{Winston} {et~al.}(2012){Winston}, {Cox}, {Prusti}, {Mer{\'{\i}}n},
  {Ribas}, {Royer}, {Vavrek}, {Puga}, {Andr{\'e}}, {Men'shchikov},
  {K{\"o}nyves}, {K{\'o}sp{\'a}l}, {Alves de Oliveira}, {Pilbratt}, \&
  {Waelkens}}]{Winston2012}
{Winston}, E., {Cox}, N.~L.~J., {Prusti}, T., {et~al.} 2012, \aap, 545, A145

\end{thebibliography}
\end{document}